\documentclass[a4paper,fleqn,usenatbib]{mnras}

\usepackage[T1]{fontenc}
\usepackage{ae,aecompl}

\usepackage{graphicx}	% Including figure files
\usepackage{amsmath}	% Advanced maths commands
\usepackage{amssymb}	% Extra maths symbols

\newcommand{\erg}{erg cm$^{-2}$ s$^{-1}$} % unit of Flux
\newcommand{\xm}{\emph{XMM-Newton}}
\newcommand{\sw}{\emph{Swift}/XRT}
\newcommand{\nus}{\emph{NuSTAR}}
\newcommand{\lxp}{\emph{AstroSat}/LAXPC}
\newcommand{\sxt}{\emph{AstroSat}/SXT}
\newcommand{\sax}{SAX J1748.9-2021}
\newcommand{\astrosat}{\emph{AstroSat}}

\title[AstroSat observation of \sax]{A broadband look of the Accreting Millisecond X-ray Pulsar SAX J1748.9-2021 using \astrosat\ and \xm}

\author[R. Sharma et al.]{Rahul Sharma$^{1}$\thanks{E-mail: rahul1607kumar@gmail.com},
Aru Beri$^{2,3}$, 
Andrea Sanna$^{4}$
and Anjan Dutta$^{1}$
\\
$^{1}$Department of Physics and Astrophysics, University of Delhi, Delhi 110007, India\\
$^{2}$DST-INSPIRE Faculty, IISER Mohali, Punjab, India 140306\\
$^{3}$School of Physics and Astronomy, University of Southampton, Southampton, Hampshire, SO17 1BJ United Kingdom \\
$^{4}$Universit\'{a} degli Studi di Cagliari, Dipartimento di Fisica, SP Monserrato-Sestu, KM 0.7, 09042 Monserrato, Italy 
}

\date{Accepted XXX. Received YYY; in original form ZZZ}

\pubyear{2020}

\begin{document}
\label{firstpage}
\pagerange{\pageref{firstpage}--\pageref{lastpage}}
\maketitle

\begin{abstract}

\sax\ is a transient accretion powered millisecond X-ray pulsar located in the Globular cluster NGC 6440.  
We report on the spectral and timing analysis of \sax\ performed on \astrosat\ data taken during its faint and short outburst of 2017. We derived the best-fitting orbital solution for the 2017 outburst and obtained an average local spin frequency of 442.361098(3) Hz. The pulse profile obtained from 3--7 keV and 7--20 keV energy bands suggest constant fractional amplitude $\sim 0.5$\% for fundamental component, contrary to previously observed energy pulse profile dependence. 
Our \astrosat\ observations revealed the source to be in a hard spectral state. The 1--50 keV spectrum from SXT and LAXPC on-board \astrosat\ can be well described with a single temperature blackbody and thermal Comptonization. Moreover, we found that the combined spectra from \xm\ (EPIC-PN) and \astrosat\ (SXT+LAXPC) indicated the presence of reflection features in the form of iron (Fe K${\alpha}$) line that we modeled with the reflection model \texttt{xillvercp}. 
One of the two X-ray burst observed during the \lxp\ observation showed hard X-ray emission ($>$30 keV) due to Compton up-scattering of thermal photons by the hot corona. Time resolved analysis performed on the bursts revealed complex evolution in emission radius of blackbody for second burst suggestive of mild photospheric radius expansion.

\end{abstract}

\begin{keywords}
accretion, accretion discs -- stars: neutron -- X-ray: binaries -- X-rays: bursts -- X-rays: individual (\sax)
\end{keywords}

%%%%%%%%%%%%%%%%%%%%%%%%%%%%%%%%%%%%%%%%%%%%%%%%%%

%%%%%%%%%%%%%%%%% BODY OF PAPER %%%%%%%%%%%%%%%%%%

\section{Introduction}

Low Mass X-ray Binaries (LMXBs) are composed of a compact object (a black hole or a neutron star) that accretes matter from a low mass companion star, $\lesssim 1M_{\sun}$. In some neutron star (NS) LMXBs, X-ray pulsations of the order of millisecond have been detected \citep[see e.g.,][]{Chakrabarty1998, Wijnands1998, Galloway2002, Markwardt2002, Papitto2013b, Sanna2018d, Sanna2018e}. These systems are called accretion powered millisecond X-ray pulsars (AMXPs) \citep[see e.g.,][for reviews]{Patruno2012, Campana2018}. The magnetic field estimated in these systems is of the order of $10^7-10^9$ Gauss \citep[see e.g.,][]{Cackett2009, Mukherjee2015, ludlam2017c, Sharma2019}. Currently, 22 AMXPs are known and all of them are transient in nature, observed during outbursts in the past 21 years \citep{Marino2019}.

\sax\ is an AMXP discovered with \emph{Beppo-SAX} during its 1998 outburst \citep{intZand1999}. It is located in the globular cluster NGC 6440 at a distance of $\sim 8.5$ kpc \citep[see e.g.,][]{Ortolani1994, Kuulkers2003, Valenti2007}.
The mass and radius of the companion star is estimated to be within the range of $0.70-0.83~M_{\odot}$ and $0.86-0.90~R_{\odot}$ \citep[see,][]{Cadelano2017}. 
Since 1998 only six outbursts have been observed in \sax\ \citep{intZand1999, intZand2001, Markwardt2005, Patruno2010, Pintore, Pintore2018, Negoro2017, Sharma2019}.  
\sax\ showed intermittent pulsations at $\sim 442.3$ Hz during its 2001, 2005, 2010 and 2015 outbursts, from which the orbital period of $\sim 8.76$ h and projected semi-major axis of $\sim 0.4$ light-seconds was inferred \citep{gavriil2007, Altamirano2008, Patruno2009, Patruno2010, Sanna2016}.

The X-ray spectrum observed during the latest outburst of 2017 revealed the presence of hard spectral state \citep{Pintore2018} similar to that observed during its 1998 outburst \citep{intZand1999}. However, during other outbursts observed in this source spectral state transition (hard to soft) was found \citep{Patruno2009,Li2018, Wu2018}. \sax\ shows the hard and soft spectral states \citep{intZand1999, Patruno2009, Pintore, Pintore2018, Li2018, Wu2018, Sharma2019} as of the atoll sources \citep{hasinger}. The spectrum of \sax\ has been described with a combination of following components: thermal emission from an accretion disc and/or NS surface, thermal comptonization and the reflected emission from the accretion disc \citep{intZand1999, Pintore, Pintore2018, Li2018, Wu2018, Sharma2019}. An additional hard power-law tail was observed during the soft state of the 2015 outburst \citep{Pintore}.

Another interesting characteristic of \sax\ is that it shows thermonuclear X-ray bursts (Type-I X-ray bursts) during its outbursts \citep{intZand1999, Kaaret2003, Galloway, Beri, Pintore, Pintore2018, Li2018, Wu2018}. 
Photospheric Radius Expansion (PRE) bursts have been observed in this source and were used to obtain mass and radius estimates of the NS \citep{guver}.

India's first dedicated multi-wavelength astronomy satellite, \astrosat\ \citep{Agrawal2006, Singh2014}, was launched in 2015.
It has five principal payloads on-board: (i) the Soft X-ray Telescope (SXT), (ii) the Large Area X-ray Proportional Counters (LAXPCs), (iii) the Cadmium-Zinc-Telluride Imager (CZTI), (iv) the Ultra-Violet Imaging Telescope (UVIT), and (v) the Scanning Sky Monitor (SSM). Here, we have performed a broadband spectroscopy using  simultaneous \xm\ and \astrosat\ (SXT and LAXPC) data of \sax\ observed during its latest outburst of 2017. 
We also report results from the timing and burst analysis carried out with \lxp\ data.

\section{Observations and data analysis}

\begin{table*}
\caption{Log of X-ray observations.}
\centering
\resizebox{2\columnwidth}{!}{
\begin{tabular}{c c c c c c c}
\hline \hline
Instrument & OBS ID & Start Time & Stop time & Mode & Exposure\\
&& (yy-mm-dd hh:mm:ss) & (yy-mm-dd hh:mm:ss) & & (ks) \\
\hline
\lxp\ & 9000001594 & 2017-10-08 07:43:55 & 2017-10-10 16:53:48 & Event Mode & 206\\
\sxt\ & 9000001594 & 2017-10-08 07:57:03 & 2017-10-10 18:40:09 & PC & 211\\
\xm/EPIC-PN & 0795712201 & 2017-10-09 11:23:17 & 2017-10-10 03:15:38  & Timing & 57\\
\hline
\end{tabular}}
\label{obslog}
\end{table*}

\begin{figure}
\centering
\includegraphics[width=1.1\linewidth]{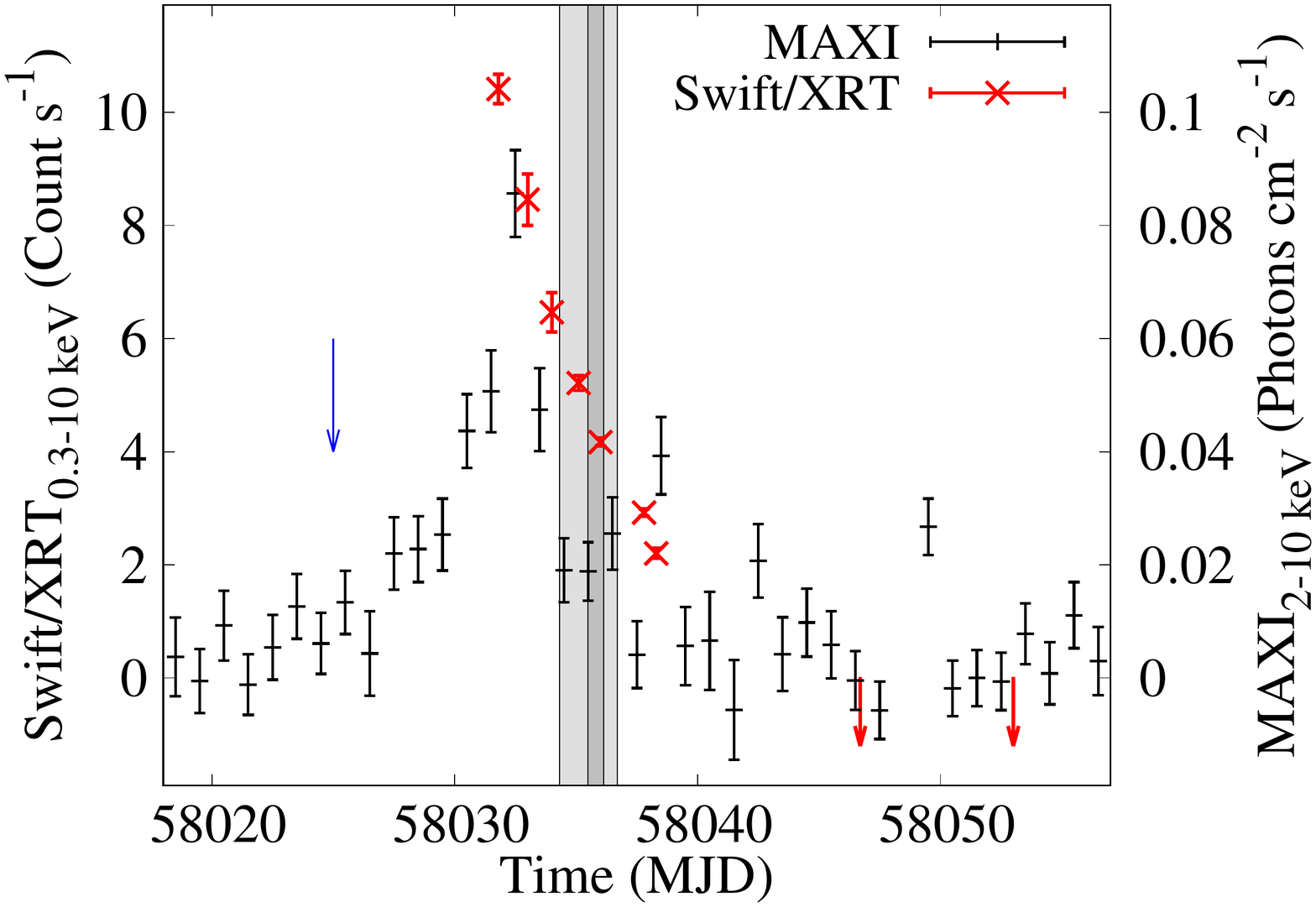}
 \caption{Timeline of \sax\ during its 2017 outburst as observed with \sw\ and \emph{MAXI}/GSC. The light grey and dark grey regions represent the time of \astrosat\ and \xm\ observations, respectively. The blue arrow corresponds to the start of 2017 outburst (MJD 58025) as reported by \citet{Negoro2017}. The first XRT observation was taken close to the peak of the outburst. The red arrows are the upper limits of the detection with XRT.}
\label{xrtlc}
\end{figure}

\subsection{\lxp}

LAXPC is one of the primary instrument aboard \emph{AstroSat}. It consists of three co-aligned identical proportional counters (LAXPC10, LAXPC20 and LAXPC30) that work in the energy range of 3--80 keV. Each LAXPC detector independently record the arrival time of each photon with a time resolution of $10 ~\mu$s and has five layers, each with 12 detector cells \citep[for details see][]{Yadav2016, Antia2017}.

Table \ref{obslog} gives the log of observations that have been used in this work.
Due to the gain instability issue caused by gas leakage, we have not used LAXPC30 data. LAXPC data were collected in the Event mode (EA) which contains the information about the time, channel number and anodeID of each event.
We have used \textsc{LaxpcSoft}\footnote{http://www.tifr.res.in/$\sim$astrosat\_laxpc/LaxpcSoft.html} software package to extract light curves and spectra. 
LAXPC detectors have dead-time of $42 ~\mu$s and the extracted products are dead-time corrected.
The background in LAXPC is estimated from the blank sky observations \citep[see][for details]{Antia2017}. We found that the source was detected up to 50 keV, therefore, to minimize the background we have performed spectroscopy using the data of top layer (L1, L2) of each detector \citep[also see][for details]{Beri2019}. We have used response files to obtain channel to energy conversion information while performing energy-resolved analysis.

We corrected the LAXPC photon arrival times to the Solar system barycentre by using the \textsc{as1bary}\footnote{http://astrosat-ssc.iucaa.in/?q=data\_and\_analysis} tool. We used the best available position of the source, R.A. (J2000)$=17^h 48^m 52.^s163$ and Dec. (J2000) $=-20^{\circ}21'32.''40$ obtained with \emph{Chandra} \citep{Pooley2002}. Timing analysis is performed on LAXPC10 and LAXPC20 data.

\subsection{\sxt}

The Soft X-ray Telescope (SXT) is a focusing X-ray telescope with CCD in the focal plane that can perform X-ray imaging and spectroscopy in the 0.3--7 keV energy range \citep{Singh2014, Singh2016, Singh2017}. 
\sax\ was observed in the Photon Counting (PC) mode with SXT (Table \ref{obslog}). Level 1 data were processed with \texttt{AS1SXTLevel2-1.4b} pipeline software to produce level 2 clean event files and these files were merged using SXT Event Merger Tool (Julia Code\footnote{http://www.tifr.res.in/$\sim$astrosat\_sxt/dataanalysis.html}). This merged event file was used to extract image, light curves and spectra using the ftool task \textsc{xselect} 2.4d, provided as part of \textsc{heasoft} version 6.22. 
A circular region with radius of 15 arcmin centered on the source was used. For spectral analysis, we have used the background spectrum (SkyBkg\_comb\_EL3p5\_Cl\_Rd16p0\_v01.pha), spectral redistribution matrix file (sxt\_pc\_mat\_g0to12.rmf) and  ancillary response file (sxt\_pc\_excl00\_v04\_20190608.arf) provided by the SXT team\footnote{http://www.tifr.res.in/$\sim$astrosat\_sxt/dataanalysis.html}.

\subsection{\xm/EPIC-PN}

\xm\ has European Photon Imaging Camera (EPIC), Reflection Grating Spectrometer (RGS) and Optical Monitor (OM) on-board. The EPIC consists of one PN camera \citep{Struder2001} and two MOS detectors \citep{Turner2001}, sensitive in the 0.1--15 keV energy range. The \xm\ observation of \sax\ has an overlap in time with the \astrosat\ observation (refer to Table \ref{obslog} for details). 
For current analysis, we have used the EPIC-PN data which was operated in the timing mode. EPIC-PN data was reduced with SAS v16.1.0 with RDPHA corrections \citep{Pintore2014}. The spectra and light curves were extracted selecting single and double pixel events with $PATTERN \leq 4$ and $FLAG=0$, which retains events optimally calibrated for spectral analysis. Following \citet{Pintore2018}, source and background events were extracted from RAWX=[32:44] and RAWX=[3:5], respectively. The spectra were rebinned with an oversample of 3 and minimum of 25 counts per bin using \textsc{specgroup} task. To avoid the EPIC-pn (timing mode) calibration uncertainties at low energies, we analyzed the spectra in 1.3--10 keV energy range \citep{Pintore2018}. 

%-------------------------------------------------------------------

\begin{figure}
\centering
\includegraphics[width=\linewidth]{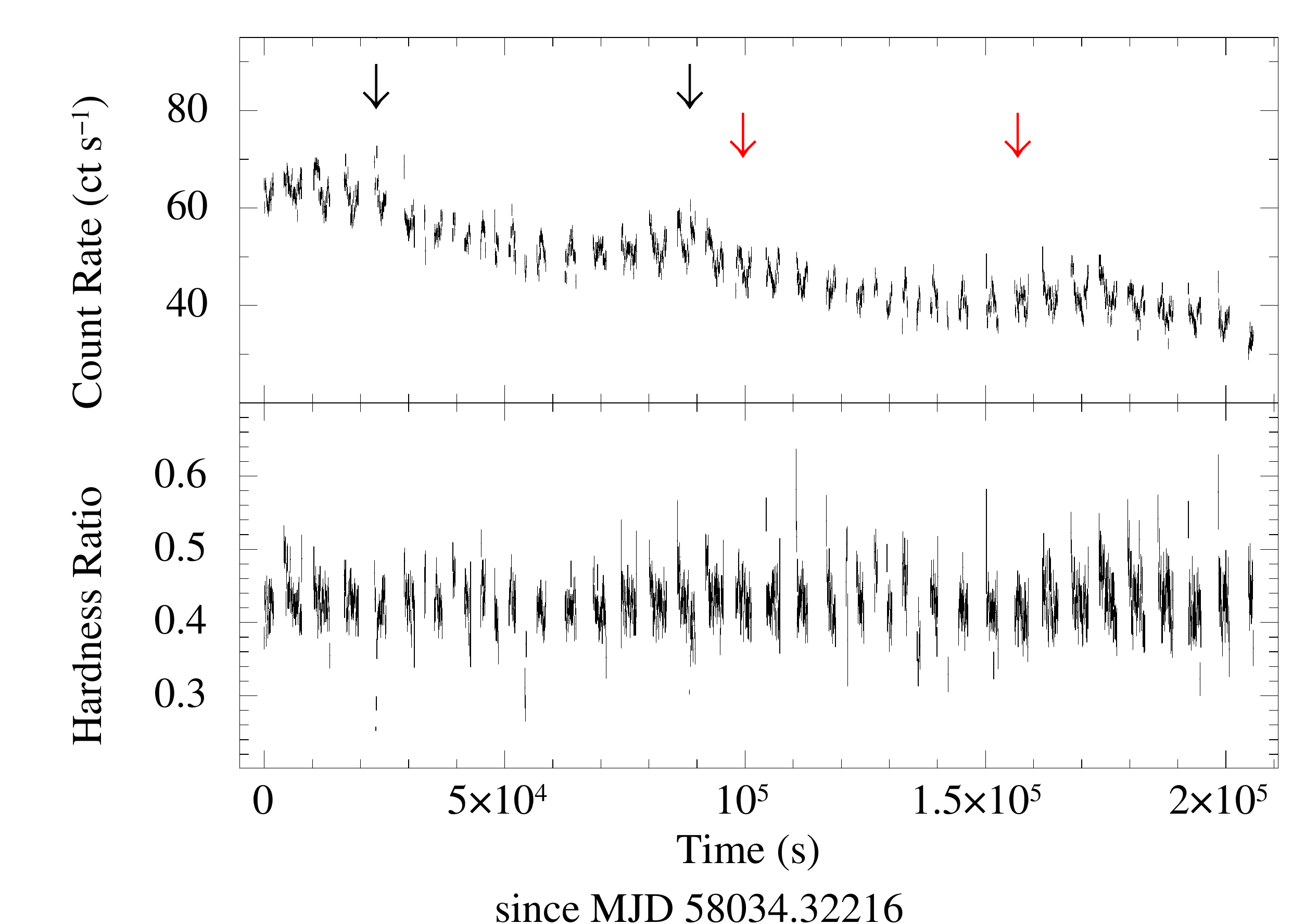}
 \caption{\textit{Top:} 3--80 keV background corrected persistent light curve from LAXPC10 binned at 100 sec and, \textit{bottom:} Hardness ratio (=hard/soft) between 3--10 keV and 10--30 keV binned at 100 sec. Red arrows indicate the times during which there is an overlap between \astrosat\ and \xm. Black arrows represents the time when X-ray bursts were observed.}
\label{hr}
\end{figure}

\section{Results}

\subsection{Light Curve}

Figure \ref{xrtlc} shows the light curve of \sax\ during its 2017 outburst as observed with the X-ray telescope (XRT)\footnote{Created from the online tool: Build XRT products \citep{Evans2007} of UK Swift Science Data Centre.} on-board  the Neil Gehrels \emph{Swift} Observatory \citep{Gehrels} and with Gas Slit Camera (GSC)\footnote{http://maxi.riken.jp/top/lc.html} on-board the Monitor of All-sky X-ray Image \citep[\emph{MAXI}; ][]{Matsuoka2009, Mihara2011}. 
Figure \ref{hr} shows the background corrected light curve extracted from LAXPC10 (upper panel) binned at 100 sec. The LAXPC light curves show the persistent emission separated by data gaps due to Earth occultation and South Atlantic Anomaly (SAA) passage. Two Type-I X-ray bursts are also observed in the LAXPC light curves (position marked with black arrows in Figure \ref{hr}).
However, these X-ray bursts were not seen in the SXT light curves (not plotted) as these times when X-ray bursts were observed have been filtered during the Good time filtering. The X-ray burst seen in the PN light curve has been reported earlier \citep{Pintore2018}. Therefore, we have excluded this X-ray burst from our analysis.

During the \astrosat\ observation, the 3--80 keV background subtracted count rate of persistent emission from LAXPC10 decreased to $\sim$half from 63 count s$^{-1}$ at the start of observation to 33 count s$^{-1}$ at the end of observation (Figure \ref{hr}). Similar trend was observed with LAXPC20 also.
However, we did not observe any change in the hardness ratio calculated using light curves in the two energy bands 3--10 keV and 10--30 keV during the observation (bottom panel of Figure \ref{hr}), suggesting that source did not seem to change the spectral state with decay in the count rate. The 0.5--10 keV count rate of \xm\ decreased to 34 count s$^{-1}$ from 38 count s$^{-1}$, during the \xm\ observation.

\subsection{Timing analysis}

We started by correcting the \lxp\ time series for the binary orbital motion \citep[see e.g.][for details on the method]{Burderi2007} through the available source ephemeris obtained during its 2015 outburst \citep{Sanna2016}. 
We then performed epoch-folding search of the whole dataset around the spin frequency value $\nu_0$ = 442.3610957 Hz (mean spin value reported during the 2015 outburst), exploring the frequency space with steps of $10^{-7}$ Hz for a total of 1001 steps. Moreover, we investigated possible updates on the orbital solution by exploring parameters with the largest propagated uncertainty with respect to the orbital solution of reference. More specifically, we focused on the time of passage from the ascending node characterised by a propagated uncertainty of $\sigma_{ T^{\star}_{2017}}\sim780$ s, obtained assuming a constant orbital period.

We corrected photon time of arrivals adopting the 2015 orbital ephemeris, except for $T^{\star}$, that we varied in steps of 1 seconds in the range $T_{2017}^{\star}\pm\sigma_{ T^{\star}_{2017}}$. Epoch-folding techniques are then applied to search for X-ray pulsation around the spin frequency $\nu_0$ using 8 phase bin to sample the signal. The most significant pulse profile has been obtained for $\Delta T^{\star}=480.6$ seconds and $\bar{\nu}=442.361098$ Hz. We note that the $T^{\star}$ value obtained is consistent within errors with the expected value propagated from the previous solution. Following \citet{Riggio2011, Sanna2016}, we estimated the uncertainty on $T^{\star}$ and $\bar{\nu}$ using Monte Carlo simulations (100 datasets to allow the $1\sigma$ error estimations), obtaining the values $T_{2017}^{\star} = T_{2015}^{\star} + \Delta T^{\star} = 58034.28452(1)$ MJD(TDB) and $\bar{\nu}=442.361098(3)$ Hz.

\begin{figure}
\centering
\includegraphics[width=0.9\columnwidth]{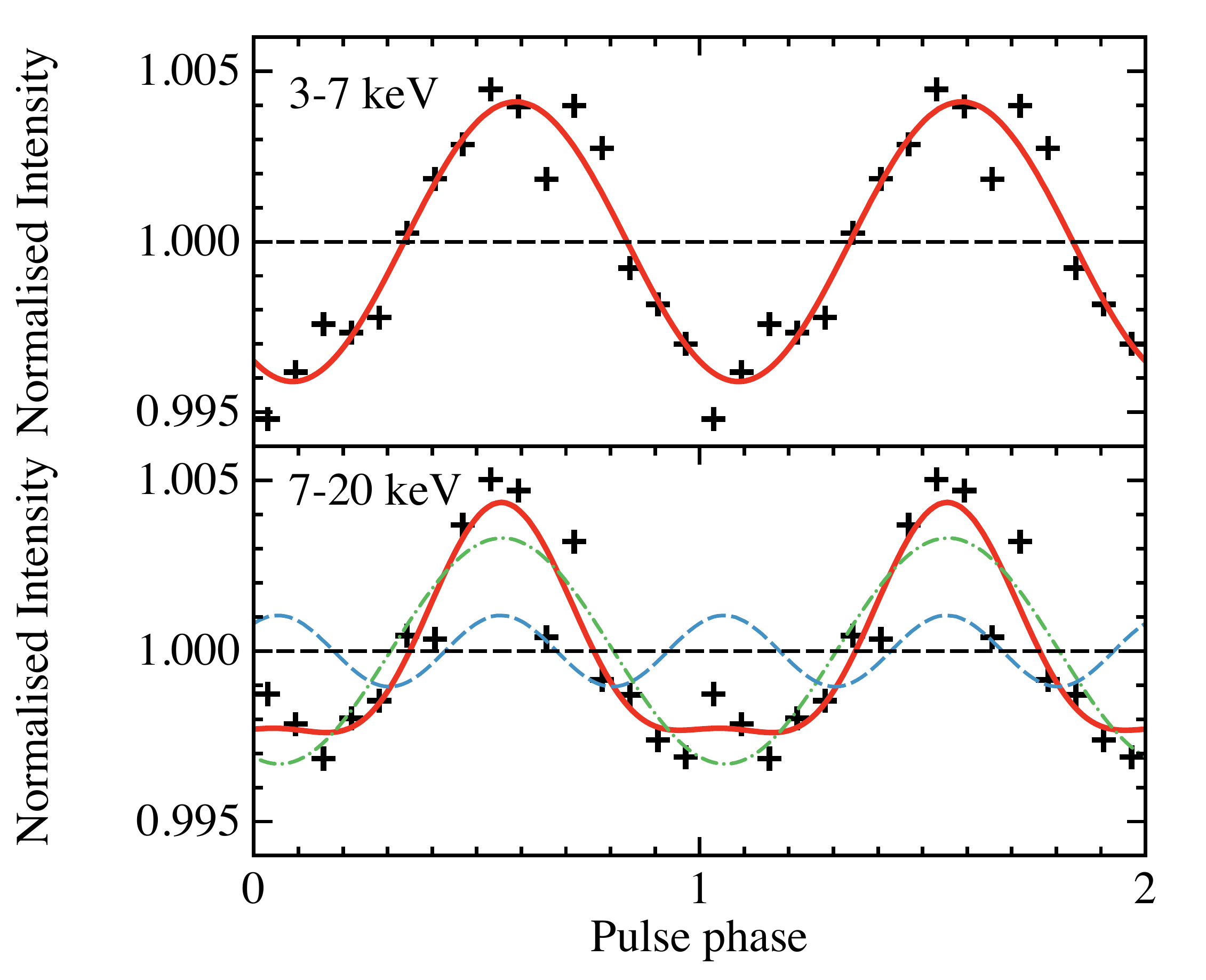}
 \caption{\sax\ pulse profiles (black points) obtained epoch-folding the \lxp\ data in the energy range 3--7 keV (top panel) and 7--20 keV (bottom panel) after correcting for the updated orbital solution. The best-fitting model (red line) is the superposition of one and two sinusoidal functions with harmonically related periods, respectively. For clarity, we show two cycles of the pulse profile.}
\label{fig:pp}
\end{figure}

Finally, in Figure \ref{fig:pp} we report the best pulse profile obtained epoch-folding the dataset at $\bar{\nu}$ and sampling the signal in 16 phase bins for the energy bands 3--7 keV (top panel) and 7--20 keV (bottom panel). The pulse shape of the 3-7 keV energy band is well fitted with a sinusoidal function with background corrected fractional amplitude of 0.5\%. The 7--20 keV pulse profile requires two harmonically related components with background corrected fractional amplitude of 0.45\% and 0.15\% for the fundamental and first overtone, respectively.

We also checked for burst oscillations during the two observed type-I bursts with LAXPC data, but no significant X-ray pulsation compatible with the spin frequency of the source seems to be present. Even after combining the two bursts, no significant pulsations were detected.

\subsection{Burst Profiles}
\begin{figure*}
\centering
\includegraphics[width=\columnwidth]{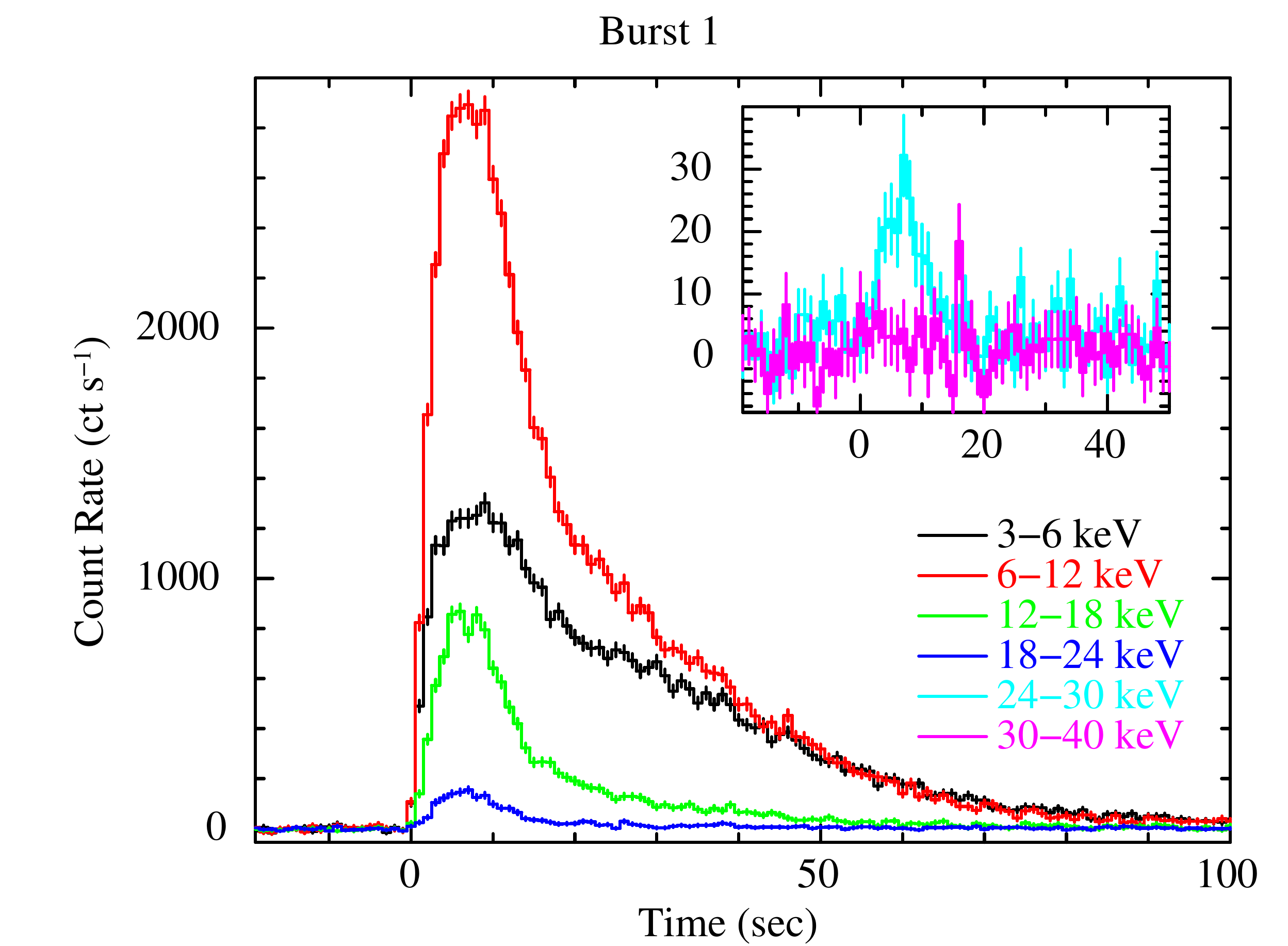}
\includegraphics[width=\columnwidth]{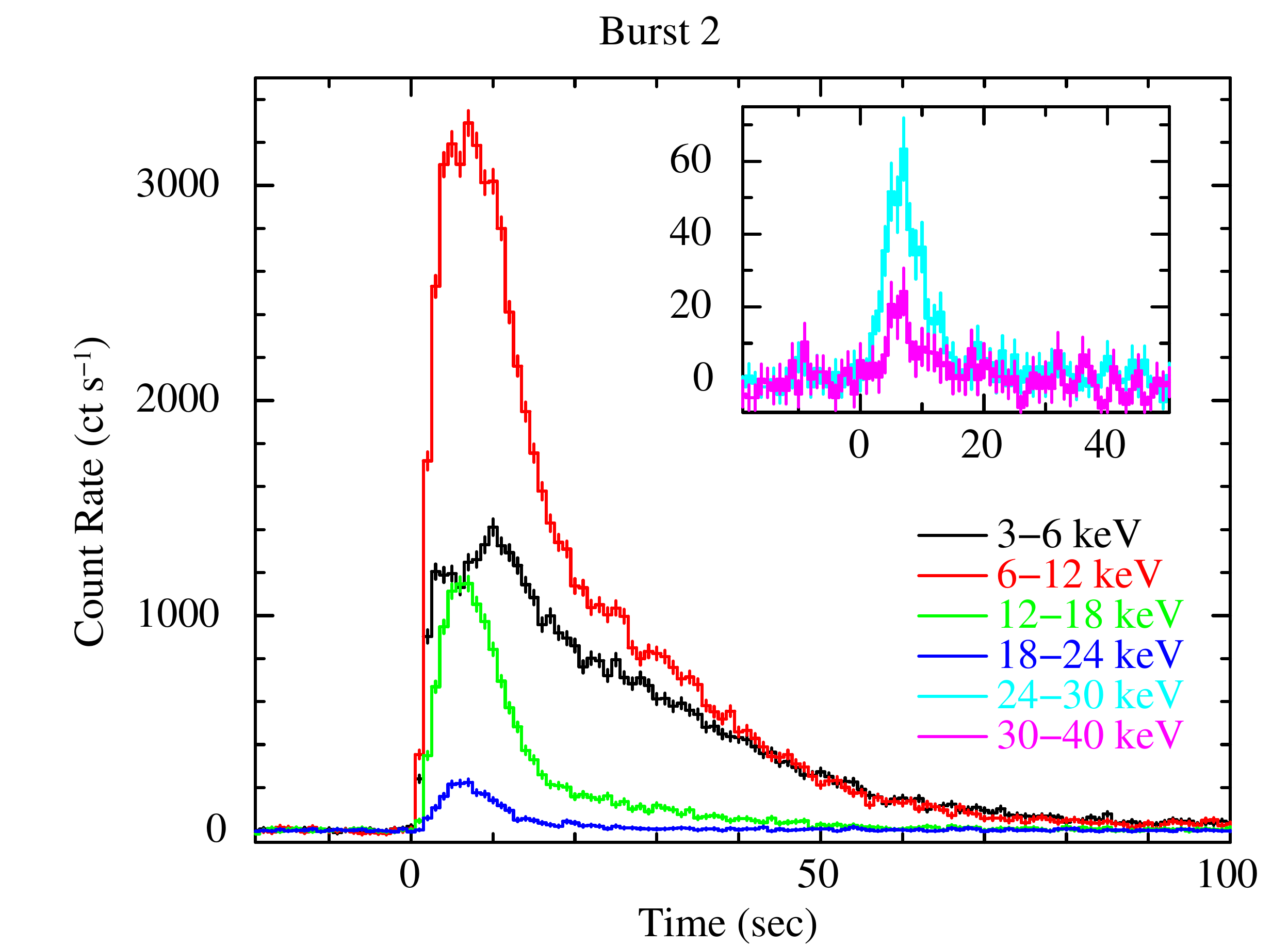}
 \caption{\astrosat-LAXPC (LAXPC10+LAXPC20) background-corrected light curve of the two X-ray bursts in different energy bands. The inset shows light curves in two energy bands 24--30 keV and 30--40 keV.}
\label{profile}
\end{figure*}

\begin{figure}
\centering
\includegraphics[width=0.9\columnwidth]{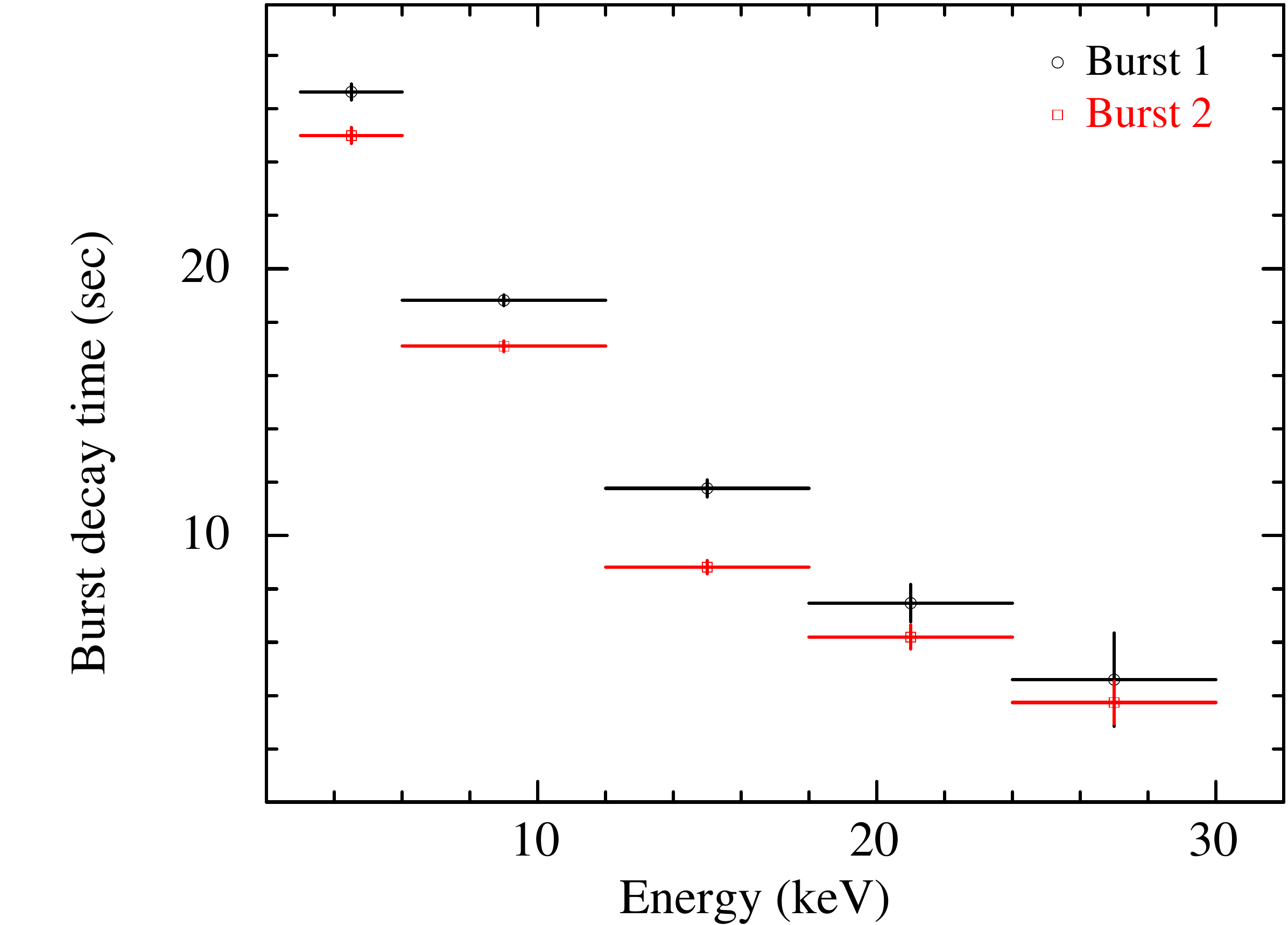}
 \caption{The exponential decay time of two bursts as a function of energy.}
\label{decay}
\end{figure}

To understand the energy dependence of X-ray bursts, we extracted light curves in different energy bands namely, 3--6 keV, 6--12 keV, 12--18 keV, 18--24 keV, 24--30 keV and 30--40 keV. Figure \ref{profile} shows  burst profiles created using the combined data of LAXPC10 and LAXPC20. Light curves are binned with a binsize of 1 sec.
\sax\ has exhibited a wide variety of burst profiles \citep{Galloway, Beri, Pintore, Pintore2018, Li2018}. To quantify the behavior of observed bursts, decay times were measured by modeling the bursts profiles using linear rise followed by an exponential decay. 
We measured the exponential decay time of both bursts in different energy band. We found that the burst duration decreases with increasing energy. 
Figure \ref{decay} shows the gradual decrease in decay time with increasing energy due to cooling of the burst to lower temperature with the decay of burst \citep{Degenaar2016b, Beri2019}. The first X-ray burst was detected up to 30 keV while the second was observed up to 40 keV (see inset in Fig. \ref{profile}). Following \citet{Beri2019}, we also checked for the presence of dips due to effect of X-ray burst on the hard X-ray emission in hard X-ray light curves during the observed bursts (30-80 keV for burst 1 and 40-80 keV for burst 2). No dip in the hard X-ray light curve was observed during any of the bursts.

\subsection{Time-Resolved Burst Spectroscopy}

To understand the spectral evolution during these X-ray bursts, we have performed time-resolved spectroscopy using spectra extracted with a duration of 1 sec. Spectra obtained from LAXPC10 and LAXPC20 were fitted simultaneously in the energy band of 3--30 keV. We added a cross-calibration constant between the two LAXPC instruments. For all burst intervals, a spectrum extracted from 90 s of data preceding the burst was extracted as the underlying accretion emission.

We fitted each spectrum with a blackbody function (\texttt{bbodyrad}) in \textsc{xspec} v 12.9.1m \citep{Arnaud}. \texttt{Tbabs} was used to model interstellar absorption with abundances set to \texttt{WILM} \citep{Wilms} and the cross-sections to \texttt{VERN} \citep{Verner}. We fixed the interstellar column density to $N_H= 0.58 \times 10^{22}$ cm$^{-2}$ \citep{Pintore}. 

\begin{figure*}
\centering
\includegraphics[width=1.03\columnwidth]{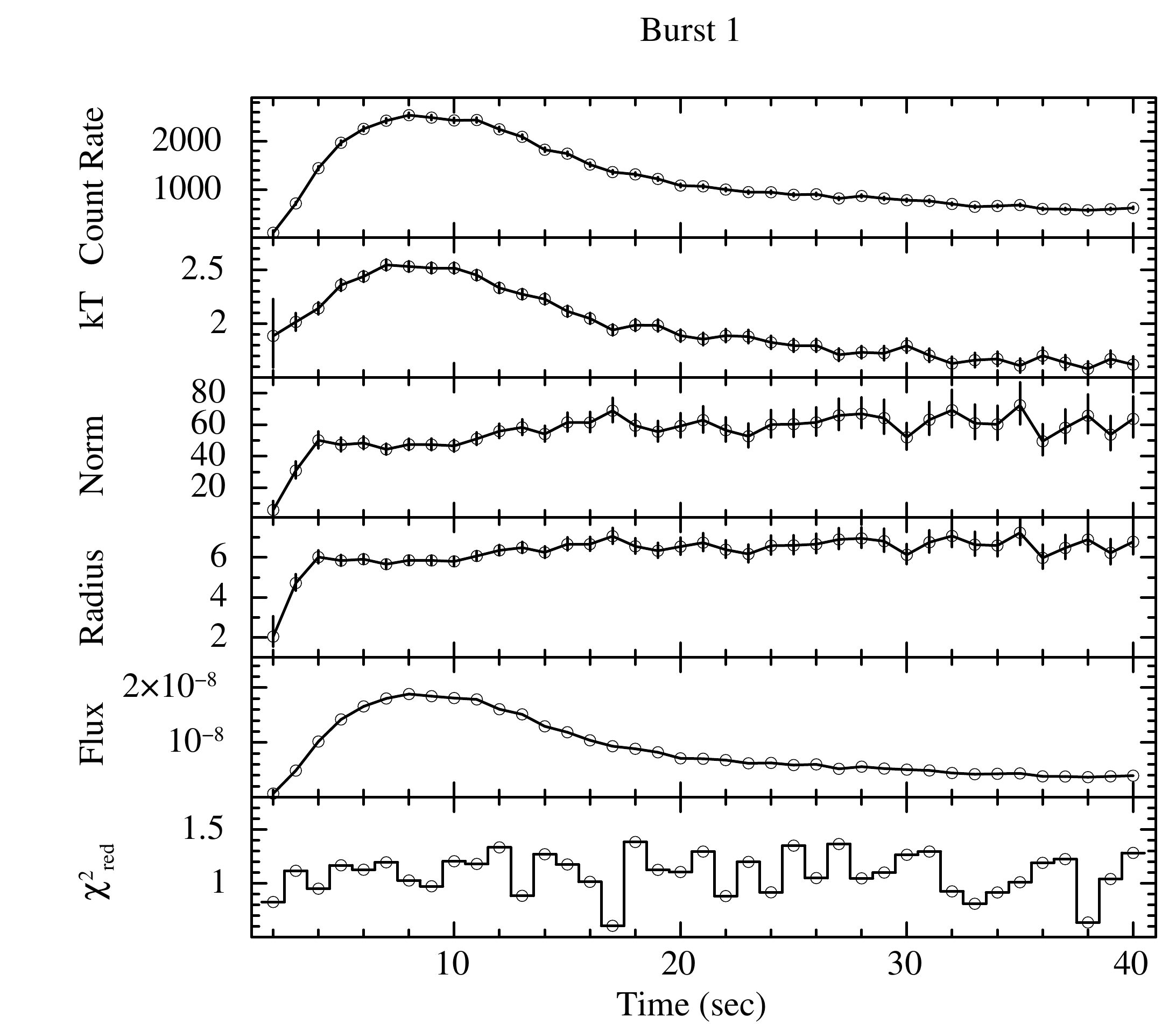}
\includegraphics[width=1.03\columnwidth]{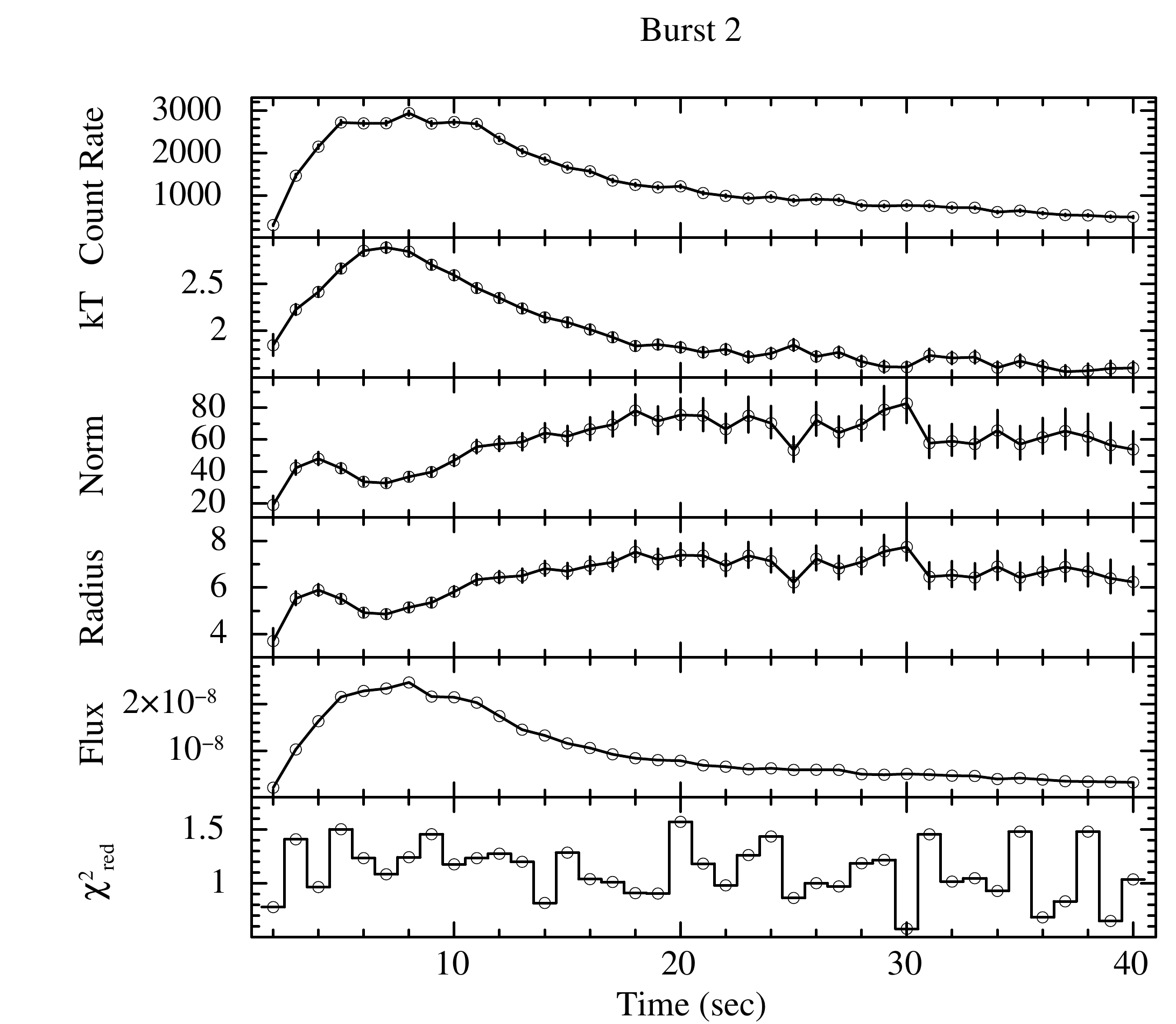}
 \caption{Time resolved spectroscopy of the two bursts observed with \lxp\ observation.}
\label{trs}
\end{figure*}

The evolution of count rate in 3--30 keV, blackbody temperature ($kT_{BB}$) in keV, blackbody normalisation ($N_{BB}$), emission radius in km, absorbed flux in 3--30 keV in units of \erg\ and reduced $\chi^2$ during each burst are plotted in Figure \ref{trs} from top to bottom, respectively.
Burst 2 was brighter than burst 1 and the temperature measured during the peak of this burst is $2.88 \pm 0.05$ keV. Moreover, the evolution of the blackbody radius indicates the presence of PRE phase. The peak temperature of burst 1 was observed to be $2.54 \pm 0.05$ keV. We calculated the bolometric flux ($F_{bol}$) using $F_{bol}=1.076~N_{BB}~(kT_{BB})^4 \times 10^{-11}$ \erg\ \citep{Galloway}. At peak, bolometric flux was observed to be $2.09 \times 10^{-8}$ \erg\ and $2.57 \times 10^{-8}$ \erg\ for burst 1 and 2, respectively.

%%%%%%%%%%%%%%%%%%%%%%%%%%%%%%%%%%%%%%%%%%%%%%%%%%%%%%%%%%%%%%

\begin{table}
\centering
\caption{The obtained best fit spectral parameters for \sax. Reported errors and limits are at 90\% for one parameter.}
\resizebox{1.05\linewidth}{!}{
\begin{tabular}{c c c c}
\hline
Model & Parameters & SXT+LAXPC & SXT+XMM+LAXPC\\
\hline
TBabs & $N_H$ ($10^{22}$ cm$^{-2}$) & $0.67 \pm 0.10$ & $0.74 \pm 0.08$ \\
\\
Bbodyrad & $kT_{BB}$ (keV) & $1.12 \pm 0.08$ & $0.60 \pm 0.04$ \\
        & Norm & $2.7^{+1.2}_{-1.0}$ & $13.2 \pm 1.5$ \\
\\
Nthcomp & $\Gamma$ & $1.60 \pm 0.04 $ & $1.719 \pm 0.025$ \\[0.5ex]
        & $kT_{seed}$ (keV) & $0.35 \pm 0.06$ & $=kT_{BB}$\\[0.5ex]
        & $kT_e$ (keV) & $>15.5$ & $>14$\\[0.5ex]
        & Norm  & $0.024^{+0.004}_{-0.003}$ & $0.012 \pm 0.003$ \\
\\
XillverCP & log $\xi$ & & $3.5 \pm 0.3$ \\[0.5ex]
          & $A_{Fe}$  &  & $1.5^{+2.7}_{-0.8}$ \\[0.5ex]
          & Norm  ($10^{-4}$) &  & $1.1^{+1.1}_{-0.7}$ \\
\\
Constant  & $C_{LAXPC}$ & 1 (fixed) & 1 (fixed) \\
          & $C_{SXT}$ & $1.12 \pm 0.04$ & $1.19 \pm 0.03$ \\
          & $C_{XMM}$ & - & $1.011 \pm 0.016$ \\
\\
Unabs. Flux & $F_{0.1-100~keV}$  & $9.5\times 10^{-10}$ & $7.73\times 10^{-10}$ \\
            & (\erg) & & \\
          & $\chi^2/dof$ & 599.8/524 & 527.5/493 \\
\hline
\end{tabular}}
\label{bb-para}
\end{table}

%------------------------------------------------------------

\subsection{Broadband Spectral Analysis}

We performed a broadband spectroscopy (1--50 keV)  using the data during persistent emission obtained with SXT (1--7 keV) and LAXPC (3--50 keV) aboard \astrosat. Data below 1 keV was ignored due to low energy calibration issue of SXT.
We used only LAXPC10 detector from LAXPC instrument for broadband spectral fitting, LAXPC20 was avoided due to instrument calibration issues at higher energies. The LAXPC10 spectra was regrouped as 0--99 by 2 channels, 100--199 by 4 channels and above 200 by 8 channels. A systematic uncertainty of 2\% was added to LAXPC spectra \citep{Antia2017, Sreehari2019}.
The SXT spectra was grouped using \textsc{grppha} to have a minimum of 25 counts per bin. While performing spectral fitting, we added a multiplicative constant component to account for  cross-calibration between two instruments. The parameter value of the constant was fixed to 1 for LAXPC10 and for SXT allowed to vary. We also allowed the gain of the response file of SXT to vary, with slope fixed to 1. We obtained a gain offset of $\sim 37$ eV. We have used \texttt{tbabs} to model interstellar neutral hydrogen absorption.

\begin{figure}
\centering
\includegraphics[width=\columnwidth]{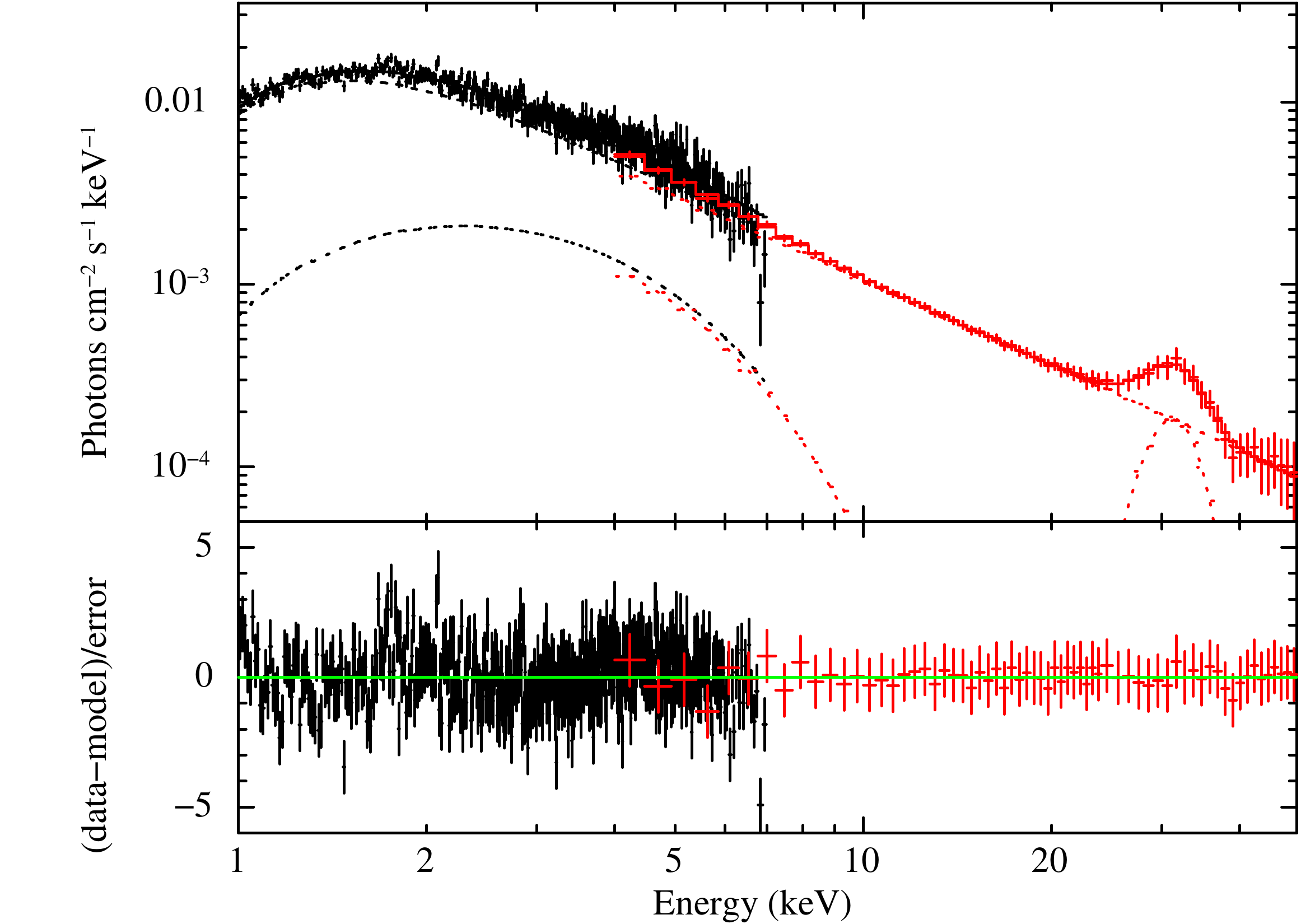}
 \caption{Best fitted time-averaged broadband spectra of SXT and LAXPC modeled with \texttt{bbodyrad+nthcomp}.}
\label{spectra}
\end{figure}

The X-ray spectral continuum of \sax\ during its 2017 outburst was best fitted using a blackbody and Comptonization model \citep{Pintore2018}. However, we noticed that during its 1998 outburst, the X-ray spectral continuum was best fitted with a single thermal Comptonized emission \citep{intZand1999}.  Therefore, we began to model the combined spectra from SXT and LAXPC using a thermal Comptonized model \texttt{nthcomp} \citep{zdziarski,zycki}.
We found that \texttt{tbabs*nthcomp} gave an unsatisfactory fit, $\chi^2$/dof = 631/526.
We also observed residuals around 30 keV which is due to the Xenon calibration edge \citep{Antia2017} and modeled using a Gaussian. Addition of a second blackbody component (soft thermal component) improved the fit and we obtained a value of $\chi^2$/dof to be 599.8/524 (Figure \ref{spectra}). We would like to mention that we did not find any residuals around 6.4 keV (due to Fe K$\alpha$) as observed with \xm\ \citep[see][]{Pintore2018}. We estimated the upper limit on the equivalent width of Fe K$\alpha$ emission line to be $\sim 20$ eV, where line energy and width was fixed to 6.5 keV and 0.2 keV, respectively. The estimated upper limit on Fe emission line is well consistent with the equivalent width ($\sim 15$ eV) measured by \citet{Pintore2018} with \xm\ data.

We report the average, unabsorbed 0.1--100 keV flux during \astrosat\ observation was $9.5\times 10^{-10}$ \erg, corresponds to unabsorbed luminosity of $L_X \sim 8.2 \times 10^{36}$ erg s$^{-1}$ for a distance of 8.5 kpc.

\subsubsection{\xm+SXT+LAXPC spectrum}
\begin{figure}
\centering
\includegraphics[width=\columnwidth]{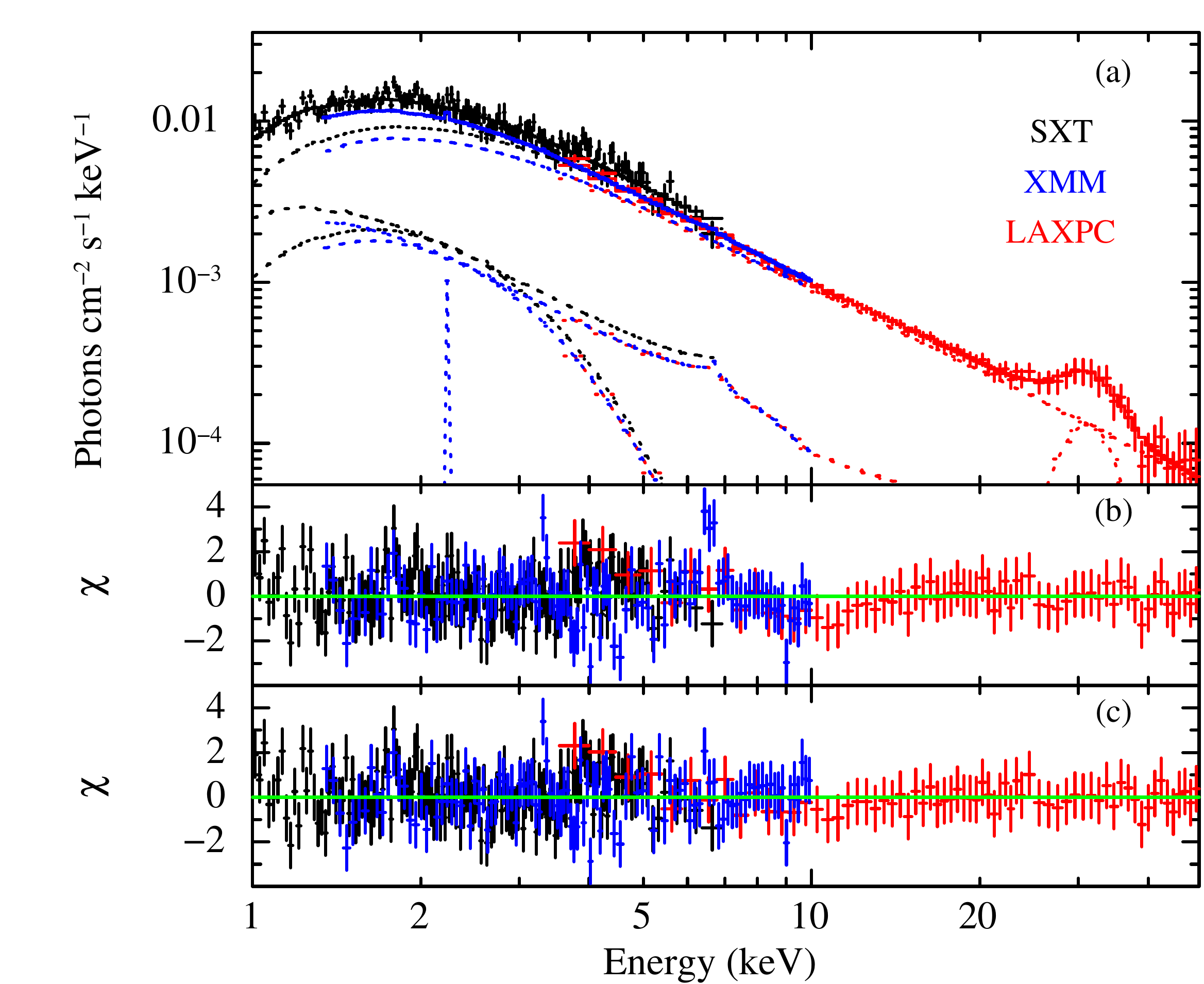}
 \caption{(a) Broadband spectra of \sax\ from \xm, SXT and LAXPC fitted simultaneously. (b) Residuals ($\chi$=(data-model)/error) with \texttt{bbodyrad+nthcomp} model. (c) Residuals with \texttt{bbodyrad+nthcomp+xillvercp} model. The figure has been rebinned for the plotting purpose.}
\label{spectra2}
\end{figure}

We extracted the SXT and LAXPC spectra using data that was strictly simultaneous with the \xm\ observation.  We removed the X-ray burst from the \xm\ data to obtain the spectrum during the persistent emission. These three spectra were simultaneously fitted with absorbed blackbody and Comptonized blackbody model.   
We fixed the value of calibration-constant to be 1 for LAXPC and we let it free to vary for SXT and EPIC-PN.
Gaussian models at $\sim$ 2.2 keV and 30 keV were included to account for the Au-M calibration edge  feature of \xm\ \citep{Papitto2009, Ferrigno2014} and Xenon calibration edge feature of LAXPC \citep{Antia2017}, respectively, as seen in the residuals.
We also found systematic residuals around 6.5 keV (see Fig-\ref{spectra2}b), arising due the Fe K emission feature. This feature has also been reported by \citet{Pintore2018}.
Therefore, we added a Gaussian model component and obtained the emission line energy at $6.43^{+0.22}_{-0.45}$ keV having a width of $0.77^{+0.75}_{-0.67}$ keV. The equivalent width of this iron line feature is about 63 eV ($\chi^2/dof=532/493$). With a fixed width of 0.13 keV, emission line energy found at $6.54 \pm 0.06$ keV with equivalent width of 18.5 eV ($\chi^2/dof=533/494$), consistent with \citet{Pintore2018}. 
Next, we added the reflection component \texttt{xillvercp} \citep{Garcia2010, garcia2013} to account for the emission line, assuming it to be originating due to accretion disc reflection. The spectral shape of the \texttt{xillvercp} was assumed to be same as \texttt{nthcomp}. The inclination angle of the accretion disc was unconstrained during the fit, so we fixed it to $32^\circ.3$ \citep{Sharma2019}. The free parameters of \texttt{xillvercp} were $log \xi$, iron abundance ($A_{Fe}$) and normalization. After adding the reflection component, fit improved to $\chi^2/dof=527.5/493$ with F-test probability of $\sim 5 \times 10^{-9}$. The disc was found to be highly ionized with ionization parameter of $\xi \sim 3200$ erg cm s$^{-1}$.
The blackbody temperature was found to be $\sim 0.6$ keV and only lower limits of $>14$ keV on the electron temperature of corona was obtained, consistent with \citet{Pintore2018}. Additionally, we found lower value of SXT gain offset $\sim 17$ eV. The best fit parameter values are presented in Table \ref{bb-para} and best fit spectrum is shown in Figure \ref{spectra2}a with residuals in \ref{spectra2}c.

We also report the average, unabsorbed flux in 0.1--100 keV energy range during \xm\ and \astrosat\ strictly simultaneous observation was $7.73\times 10^{-10}$ \erg. As a note to mention, \citet{Pintore2018} reported under-estimated fluxes in 0.3--70 keV energy range by a factor of $\sim 2$, which could be due to upper bound on the response matrix of \xm\ at 16 keV. Using the same parameter value reported in \citet{Pintore2018}, we found the 0.3--70 keV unabsorbed flux of $7.6\times 10^{-10}$ \erg\ for 9th October (XMM+ISGRI) observation, consistent with XMM+SXT+LAXPC observation and $4.4\times 10^{-10}$ \erg\ for 11th October (XRT+NUSTAR) observation.

%%%%%%%%%%%%%%%%%%%%%%%%%%%%%%%%%%%%%%%%%%%%%%%%%%%%%%%%%%%%%%

\section{Discussion and Conclusion}

\sax\ was observed in its sixth outburst in 2017. The \astrosat\ observed it for $\sim 2.5$ days, where the average LAXPC10 count rate decreased to 33 counts s$^{-1}$ from 63 counts s$^{-1}$. The LAXPC light curves also showed the presence of two type-I thermonuclear X-ray bursts.

The 1--50 keV X-ray continuum of \sax\ can be well described with the blackbody and Comptonized blackbody. The best-fit spectral parameters showed that the source was in the hard spectral state during the \astrosat\ observations (Table \ref{bb-para}). AMXPs show the hard spectrum with electron temperature of 30--50 keV \citep{Falanga2005, gierlinski2005, Papitto2010, Papitto2013, Wilkinson2011, Sanna2018a, Sanna2018b, Salvo2019}. Also, it was in hard state during the \nus\ observation taken after 1 day of \astrosat\ observation (on 2017-10-11), where the source flux reduced by 40\% \citep{Pintore2018}. 
Thus, it seems that the 2017 outburst of \sax\ did not show any spectral state change as observed during its previous outbursts in 2001, 2005, 2010 and 2015 \citep{Patruno2009,Li2018,Wu2018}. Previously, \sax\ was found to be only in the hard state during 1998 outburst \citep{intZand1999}. However, we see that the 1998 outburst of \sax\ also showed a similar behaviour, where peak luminosity reached $\sim 10^{37}$ erg s$^{-1}$ and the outburst lasted for $\sim 10$ days \citep{intZand1999, Altamirano2008}.  

We also found that the combined spectra from EPIC-PN, SXT and LAXPC showed a best fit on adding the self-consistent reflection component to the model. The obtained value of the ionization parameter indicated the presence of strongly ionized accretion disc, $\xi \sim 3200$ erg cm s$^{-1}$ \citep[e.g.,][]{Papitto2010, Salvo2019}. The iron abundance, $A_{Fe}=1.5^{+2.7}_{-0.8}$, obtained from the fit is consistent with Solar values, although the uncertainty is large. The Comptonized emission associated to a hot corona or the accretion column is characterized by a photon index of $1.719 \pm 0.025$. The electron temperature was unconstrained, only lower limit of 14 keV was obtained. A thermal blackbody with temperature of $0.60 \pm 0.04$ keV and emission radius of $\sim 3$ km, likely originating from the neutron star surface is found.

X-ray coherent pulsations at $\sim442$ Hz are significantly detected in the \astrosat\ dataset. Timing analysis of the collected events allowed us to obtain an updated orbital solution of the source, compatible within the errors with the solution obtained for the 2015 outburst of \sax\ \citep{Sanna2016}. No X-ray pulsation has been detected on timescales shorter than the whole observation. The strength of the X-ray pulsation did not allow a detailed study of the signal as a function of energy. However, the pulse profiles obtained in the energy bands 3--7 keV and 7--20 keV suggest a constant fractional amplitude around $0.5$\% for the fundamental component. This result is in contrast with the pulse profile energy dependence reported for the previous outbursts \citep[see e.g.][]{Patruno2009, Sanna2016} where the fractional amplitude has been observed to increase from 0.1\% at 0.5 keV to 4\% at 20 keV.  

In AMXPs, accretion taking place on the NS is guided by the magnetic field of the NS. This magnetically channeled accretion means that the accretion disc radius is outside the NS surface and smaller than the co-rotation radius \citep{Pringle1972, Illarionov1975}. \citet{Mukherjee2015} estimated the upper limits on magnetic field strength of 14 AMXPs, by assuming that the inner edge of the disc can not be outside the co-rotation radius. As X-ray pulsations has been observed during \astrosat\ observation, imply on going magnetically channelled accretion on to the NS. At the lowest X-ray luminosity (accretion rate), the accretion disc cannot be outside the co-rotation radius.
Using the above assumptions, we estimated the upper limit on the magnetic field strength of NS from the flux obtained with XMM+SXT+LAXPC data ($F_{min}=7.73 \times 10^{-10}$ \erg). Using equation (10) of \citet{Mukherjee2015}, the upper limit on magnetic dipole moment estimated to be $1.6 \times 10^{27}$ G cm$^3$, consistent with the estimates of \citet{Sharma2019}. This will give $B<1.6 \times 10^9$ G for \sax\  which is nearly a factor of 2 lower than the previous estimate of \citet{Mukherjee2015}.

\subsection{Burst Analysis}

We performed the time-resolved burst spectroscopy and energy-resolved burst analysis on the two X-ray bursts observed with \lxp. 
From the time-resolved spectroscopy performed on the 1 sec bin of burst, a complex variation in emission radii of blackbody was found in burst 2, suggestive of mild PRE phase. The second burst was brighter than the first one. 
The burst 1 was detected upto 30 keV only, but burst 2 showed emission in 30--40 keV energy range also. The burst observed with \emph{Beppo-SAX} during 1998 also showed the emission $>30$ keV, where the burst emission is Compton up-scattered by the hot corona \citep{intZand1999}. The bursts were also found to influence the hard to soft state transition time. The soft photons from the bursts cool the corona faster to push the state transition to the shorter time scale \citep{Li2018}. The burst decay time strongly depends on the energy and the decay time of burst 2 was lower than the burst 1.
The peak temperature and bolometric flux of $2.88$ keV and $2.57 \times 10^{-8}$ \erg, respectively were obtained for burst 2. Previously, observed PRE bursts of \sax\ showed the peak flux of $\sim 2.8-4 \times 10^{-8}$ \erg \citep{Galloway}.

The local accretion rate per unit area onto the compact object can be estimated using $\dot{m} = L_{pers} (1+z) ((4\pi R^2)(GM/R))^{-1}$. Using the gravitational redshift of $1+z=1.31$ for a canonical NS with a mass $M=1.4M_{\sun}$ and a radius of $R=10$ km, we found $\dot{m} \simeq 0.48 \times 10^4$ g cm$^{-2}$ s$^{-1}$. The observed recurrence time depends on $\dot{m}$ as $\Delta t_{rec} \propto <\dot{m}>^{-1.02}$ measured from the burst observed during 2015 outburst \citep{Li2018}. From the estimated average accretion rate of 2017 outburst, the  burst recurrence time estimated was $\sim 6$ h. The shortest difference between the observed burst of 2017 was $\sim 8$ h, consistent with above estimation. The three burst observed with wide field camera of \emph{BeppoSAX} in 1998 outburst showed the recurrence time of $\simeq 2.8$ h \citep{intZand1999}.

\section*{Acknowledgements}

RS acknowledges the financial support from the University Grants Commission (UGC), India, under the Senior Research Fellow (SRF) scheme. AB is grateful to both the Royal Society, U.K and to SERB (Science and Engineering Research Board), India. AB is supported by an INSPIRE Faculty grant (DST/INSPIRE/04/2018/001265) by the Department of Science and Technology, Govt. of India and also acknowledges the financial support of Indian Space Research Organisation (ISRO) under \astrosat\ archival Data utilization program. AB would also like to thank Prof. K. P. Singh for several discussions regarding SXT data analysis.   
This publication uses data from the \astrosat\ mission of ISRO, archived at the Indian Space Science Data Centre (ISSDC) and \xm\ for which the data was obtained from High Energy Astrophysics Science Archive Research Center (HEASARC), provided by NASA's Goddard Space Flight Center. We thank the LAXPC Payload Operation Center (POC) and the SXT POC at TIFR, Mumbai for providing necessary software tools. We would like to thank IISER, Mohali for extending its library facilities. The authors  thank  the  anonymous  referee  for  the valuable comments on the manuscript.

%%%%%%%%%%%%%%%%%%%%%%%%%%%%%%%%%%%%%%%%%%%%%%%%%%



%\begin{thebibliography}{99}
%\end{thebibliography}

%%%%%%%%%%%%%%%%%%%%%%%%%%%%%%%%%%%%%%%%%%%%%%%%%%

%%%%%%%%%%%%%%%%% APPENDICES %%%%%%%%%%%%%%%%%%%%%

%\appendix


%%%%%%%%%%%%%%%%%%%%%%%%%%%%%%%%%%%%%%%%%%%%%%%%%%

% Don't change these lines
\bsp	% typesetting comment
\label{lastpage}

\begin{thebibliography}{}
\makeatletter
\relax
\def\mn@urlcharsother{\let\do\@makeother \do\$\do\&\do\#\do\^\do\_\do\%\do\~}
\def\mn@doi{\begingroup\mn@urlcharsother \@ifnextchar [ {\mn@doi@}
  {\mn@doi@[]}}
\def\mn@doi@[#1]#2{\def\@tempa{#1}\ifx\@tempa\@empty \href
  {http://dx.doi.org/#2} {doi:#2}\else \href {http://dx.doi.org/#2} {#1}\fi
  \endgroup}
\def\mn@eprint#1#2{\mn@eprint@#1:#2::\@nil}
\def\mn@eprint@arXiv#1{\href {http://arxiv.org/abs/#1} {{\tt arXiv:#1}}}
\def\mn@eprint@dblp#1{\href {http://dblp.uni-trier.de/rec/bibtex/#1.xml}
  {dblp:#1}}
\def\mn@eprint@#1:#2:#3:#4\@nil{\def\@tempa {#1}\def\@tempb {#2}\def\@tempc
  {#3}\ifx \@tempc \@empty \let \@tempc \@tempb \let \@tempb \@tempa \fi \ifx
  \@tempb \@empty \def\@tempb {arXiv}\fi \@ifundefined
  {mn@eprint@\@tempb}{\@tempb:\@tempc}{\expandafter \expandafter \csname
  mn@eprint@\@tempb\endcsname \expandafter{\@tempc}}}

\bibitem[\protect\citeauthoryear{{Agrawal}}{{Agrawal}}{2006}]{Agrawal2006}
{Agrawal} P.~C.,  2006, \mn@doi [Advances in Space Research]
  {10.1016/j.asr.2006.03.038}, \href
  {https://ui.adsabs.harvard.edu/abs/2006AdSpR..38.2989A} {38, 2989}

\bibitem[\protect\citeauthoryear{{Altamirano}, {Casella}, {Patruno}, {Wijnands}
   \& {van der Klis}}{{Altamirano} et~al.}{2008}]{Altamirano2008}
{Altamirano} D.,  {Casella} P.,  {Patruno} A.,  {Wijnands} R.,   {van der Klis}
  M.,  2008, \mn@doi [\apjl] {10.1086/528983}, \href
  {http://adsabs.harvard.edu/abs/2008ApJ...674L..45A} {674, L45}

\bibitem[\protect\citeauthoryear{{Antia} et~al.,}{{Antia}
  et~al.}{2017}]{Antia2017}
{Antia} H.~M.,  et~al., 2017, \mn@doi [\apjs] {10.3847/1538-4365/aa7a0e}, \href
  {https://ui.adsabs.harvard.edu/abs/2017ApJS..231...10A} {231, 10}

\bibitem[\protect\citeauthoryear{{Arnaud}}{{Arnaud}}{1996}]{Arnaud}
{Arnaud} K.~A.,  1996, in {Jacoby} G.~H.,  {Barnes} J.,  eds,  Astronomical
  Society of the Pacific Conference Series Vol. 101, Astronomical Data Analysis
  Software and Systems V. p.~17

\bibitem[\protect\citeauthoryear{{Beri}, {Paul}, {Orlandini}  \&
  {Maitra}}{{Beri} et~al.}{2016}]{Beri}
{Beri} A.,  {Paul} B.,  {Orlandini} M.,   {Maitra} C.,  2016, \mn@doi [\na]
  {10.1016/j.newast.2015.10.013}, \href
  {http://adsabs.harvard.edu/abs/2016NewA...45...48B} {45, 48}

\bibitem[\protect\citeauthoryear{{Beri} et~al.,}{{Beri}
  et~al.}{2019}]{Beri2019}
{Beri} A.,  et~al., 2019, \mn@doi [\mnras] {10.1093/mnras/sty2975}, \href
  {https://ui.adsabs.harvard.edu/abs/2019MNRAS.482.4397B} {482, 4397}

\bibitem[\protect\citeauthoryear{{Burderi} et~al.,}{{Burderi}
  et~al.}{2007}]{Burderi2007}
{Burderi} L.,  et~al., 2007, \mn@doi [\apj] {10.1086/510659}, \href
  {https://ui.adsabs.harvard.edu/abs/2007ApJ...657..961B} {657, 961}

\bibitem[\protect\citeauthoryear{{Cackett}, {Altamirano}, {Patruno}, {Miller},
  {Reynolds}, {Linares}  \& {Wijnands}}{{Cackett} et~al.}{2009}]{Cackett2009}
{Cackett} E.~M.,  {Altamirano} D.,  {Patruno} A.,  {Miller} J.~M.,  {Reynolds}
  M.,  {Linares} M.,   {Wijnands} R.,  2009, \mn@doi [\apjl]
  {10.1088/0004-637X/694/1/L21}, \href
  {http://adsabs.harvard.edu/abs/2009ApJ...694L..21C} {694, L21}

\bibitem[\protect\citeauthoryear{{Cadelano}, {Pallanca}, {Ferraro},
  {Dalessandro}, {Lanzoni}  \& {Patruno}}{{Cadelano}
  et~al.}{2017}]{Cadelano2017}
{Cadelano} M.,  {Pallanca} C.,  {Ferraro} F.~R.,  {Dalessandro} E.,  {Lanzoni}
  B.,   {Patruno} A.,  2017, \mn@doi [\apj] {10.3847/1538-4357/aa7b7f}, \href
  {http://adsabs.harvard.edu/abs/2017ApJ...844...53C} {844, 53}

\bibitem[\protect\citeauthoryear{{Campana} \& {Di Salvo}}{{Campana} \& {Di
  Salvo}}{2018}]{Campana2018}
{Campana} S.,  {Di Salvo} T.,  2018, in {Rezzolla} L.,  {Pizzochero} P.,
  {Jones} D.~I.,  {Rea} N.,   {Vida{\~n}a} I.,  eds,  Astrophysics and Space
  Science Library Vol. 457, Astrophysics and Space Science Library. p.~149
  (\mn@eprint {arXiv} {1804.03422}), \mn@doi{10.1007/978-3-319-97616-7_4}

\bibitem[\protect\citeauthoryear{{Chakrabarty} \& {Morgan}}{{Chakrabarty} \&
  {Morgan}}{1998}]{Chakrabarty1998}
{Chakrabarty} D.,  {Morgan} E.~H.,  1998, \mn@doi [\nat] {10.1038/28561}, \href
  {https://ui.adsabs.harvard.edu/abs/1998Natur.394..346C} {394, 346}

\bibitem[\protect\citeauthoryear{{Degenaar}, {Koljonen}, {Chakrabarty}, {Kara},
  {Altamirano}, {Miller}  \& {Fabian}}{{Degenaar} et~al.}{2016}]{Degenaar2016b}
{Degenaar} N.,  {Koljonen} K.~I.~I.,  {Chakrabarty} D.,  {Kara} E.,
  {Altamirano} D.,  {Miller} J.~M.,   {Fabian} A.~C.,  2016, \mn@doi [\mnras]
  {10.1093/mnras/stv2965}, \href
  {https://ui.adsabs.harvard.edu/abs/2016MNRAS.456.4256D} {456, 4256}

\bibitem[\protect\citeauthoryear{{Di Salvo}, {Sanna}, {Burderi}, {Papitto},
  {Iaria}, {Gambino}  \& {Riggio}}{{Di Salvo} et~al.}{2019}]{Salvo2019}
{Di Salvo} T.,  {Sanna} A.,  {Burderi} L.,  {Papitto} A.,  {Iaria} R.,
  {Gambino} A.~F.,   {Riggio} A.,  2019, \mn@doi [\mnras]
  {10.1093/mnras/sty2974}, \href
  {https://ui.adsabs.harvard.edu/abs/2019MNRAS.483..767D} {483, 767}

\bibitem[\protect\citeauthoryear{{Evans} et~al.,}{{Evans}
  et~al.}{2007}]{Evans2007}
{Evans} P.~A.,  et~al., 2007, \mn@doi [\aap] {10.1051/0004-6361:20077530},
  \href {https://ui.adsabs.harvard.edu/abs/2007A%26A...469..379E} {469, 379}

\bibitem[\protect\citeauthoryear{{Falanga} et~al.,}{{Falanga}
  et~al.}{2005}]{Falanga2005}
{Falanga} M.,  et~al., 2005, \mn@doi [\aap] {10.1051/0004-6361:20053472}, \href
  {http://adsabs.harvard.edu/abs/2005A%26A...444...15F} {444, 15}

\bibitem[\protect\citeauthoryear{{Ferrigno} et~al.,}{{Ferrigno}
  et~al.}{2014}]{Ferrigno2014}
{Ferrigno} C.,  et~al., 2014, \mn@doi [\aap] {10.1051/0004-6361/201322904},
  \href {https://ui.adsabs.harvard.edu/abs/2014A&A...567A..77F} {567, A77}

\bibitem[\protect\citeauthoryear{{Galloway}, {Chakrabarty}, {Morgan}  \&
  {Remillard}}{{Galloway} et~al.}{2002}]{Galloway2002}
{Galloway} D.~K.,  {Chakrabarty} D.,  {Morgan} E.~H.,   {Remillard} R.~A.,
  2002, \mn@doi [\apjl] {10.1086/343841}, \href
  {https://ui.adsabs.harvard.edu/abs/2002ApJ...576L.137G} {576, L137}

\bibitem[\protect\citeauthoryear{{Galloway}, {Muno}, {Hartman}, {Psaltis}  \&
  {Chakrabarty}}{{Galloway} et~al.}{2008}]{Galloway}
{Galloway} D.~K.,  {Muno} M.~P.,  {Hartman} J.~M.,  {Psaltis} D.,
  {Chakrabarty} D.,  2008, \mn@doi [\apjs] {10.1086/592044}, \href
  {http://adsabs.harvard.edu/abs/2008ApJS..179..360G} {179, 360}

\bibitem[\protect\citeauthoryear{{Garc{\'\i}a} \& {Kallman}}{{Garc{\'\i}a} \&
  {Kallman}}{2010}]{Garcia2010}
{Garc{\'\i}a} J.,  {Kallman} T.~R.,  2010, \mn@doi [\apj]
  {10.1088/0004-637X/718/2/695}, \href
  {https://ui.adsabs.harvard.edu/#abs/2010ApJ...718..695G} {718, 695}

\bibitem[\protect\citeauthoryear{{Garc{\'\i}a}, {Dauser}, {Reynolds},
  {Kallman}, {McClintock}, {Wilms}  \& {Eikmann}}{{Garc{\'\i}a}
  et~al.}{2013}]{garcia2013}
{Garc{\'\i}a} J.,  {Dauser} T.,  {Reynolds} C.~S.,  {Kallman} T.~R.,
  {McClintock} J.~E.,  {Wilms} J.,   {Eikmann} W.,  2013, \mn@doi [\apj]
  {10.1088/0004-637X/768/2/146}, \href
  {https://ui.adsabs.harvard.edu/#abs/2013ApJ...768..146G} {768, 146}

\bibitem[\protect\citeauthoryear{{Gavriil}, {Strohmayer}, {Swank}  \&
  {Markwardt}}{{Gavriil} et~al.}{2007}]{gavriil2007}
{Gavriil} F.~P.,  {Strohmayer} T.~E.,  {Swank} J.~H.,   {Markwardt} C.~B.,
  2007, \mn@doi [\apjl] {10.1086/523758}, \href
  {http://adsabs.harvard.edu/abs/2007ApJ...669L..29G} {669, L29}

\bibitem[\protect\citeauthoryear{{Gehrels} et~al.,}{{Gehrels}
  et~al.}{2004}]{Gehrels}
{Gehrels} N.,  et~al., 2004, \mn@doi [\apj] {10.1086/422091}, \href
  {http://adsabs.harvard.edu/abs/2004ApJ...611.1005G} {611, 1005}

\bibitem[\protect\citeauthoryear{{Gierli{\'n}ski} \&
  {Poutanen}}{{Gierli{\'n}ski} \& {Poutanen}}{2005}]{gierlinski2005}
{Gierli{\'n}ski} M.,  {Poutanen} J.,  2005, \mn@doi [\mnras]
  {10.1111/j.1365-2966.2005.09004.x}, \href
  {http://adsabs.harvard.edu/abs/2005MNRAS.359.1261G} {359, 1261}

\bibitem[\protect\citeauthoryear{{G{\"u}ver} \& {{\"O}zel}}{{G{\"u}ver} \&
  {{\"O}zel}}{2013}]{guver}
{G{\"u}ver} T.,  {{\"O}zel} F.,  2013, \mn@doi [\apjl]
  {10.1088/2041-8205/765/1/L1}, \href
  {http://adsabs.harvard.edu/abs/2013ApJ...765L...1G} {765, L1}

\bibitem[\protect\citeauthoryear{{Hasinger} \& {van der Klis}}{{Hasinger} \&
  {van der Klis}}{1989}]{hasinger}
{Hasinger} G.,  {van der Klis} M.,  1989, \aap, \href
  {http://adsabs.harvard.edu/abs/1989A%26A...225...79H} {225, 79}

\bibitem[\protect\citeauthoryear{{Illarionov} \& {Sunyaev}}{{Illarionov} \&
  {Sunyaev}}{1975}]{Illarionov1975}
{Illarionov} A.~F.,  {Sunyaev} R.~A.,  1975, \aap, \href
  {https://ui.adsabs.harvard.edu/abs/1975A&A....39..185I} {39, 185}

\bibitem[\protect\citeauthoryear{{Kaaret}, {in 't Zand}, {Heise}  \&
  {Tomsick}}{{Kaaret} et~al.}{2003}]{Kaaret2003}
{Kaaret} P.,  {in 't Zand} J.~J.~M.,  {Heise} J.,   {Tomsick} J.~A.,  2003,
  \mn@doi [\apj] {10.1086/375053}, \href
  {https://ui.adsabs.harvard.edu/abs/2003ApJ...598..481K} {598, 481}

\bibitem[\protect\citeauthoryear{{Kuulkers}, {den Hartog}, {in't Zand},
  {Verbunt}, {Harris}  \& {Cocchi}}{{Kuulkers} et~al.}{2003}]{Kuulkers2003}
{Kuulkers} E.,  {den Hartog} P.~R.,  {in't Zand} J.~J.~M.,  {Verbunt} F.~W.~M.,
   {Harris} W.~E.,   {Cocchi} M.,  2003, \mn@doi [\aap]
  {10.1051/0004-6361:20021781}, \href
  {http://adsabs.harvard.edu/abs/2003A%26A...399..663K} {399, 663}

\bibitem[\protect\citeauthoryear{{Li} et~al.,}{{Li} et~al.}{2018}]{Li2018}
{Li} Z.,  et~al., 2018, \mn@doi [\aap] {10.1051/0004-6361/201833857}, \href
  {https://ui.adsabs.harvard.edu/abs/2018A&A...620A.114L} {620, A114}

\bibitem[\protect\citeauthoryear{{Ludlam}, {Miller}, {Degenaar}, {Sanna},
  {Cackett}, {Altamirano}  \& {King}}{{Ludlam} et~al.}{2017}]{ludlam2017c}
{Ludlam} R.~M.,  {Miller} J.~M.,  {Degenaar} N.,  {Sanna} A.,  {Cackett} E.~M.,
   {Altamirano} D.,   {King} A.~L.,  2017, \mn@doi [\apj]
  {10.3847/1538-4357/aa8b1b}, \href
  {http://adsabs.harvard.edu/abs/2017ApJ...847..135L} {847, 135}

\bibitem[\protect\citeauthoryear{{Marino} et~al.,}{{Marino}
  et~al.}{2019}]{Marino2019}
{Marino} A.,  et~al., 2019, \mn@doi [\aap] {10.1051/0004-6361/201834460}, \href
  {https://ui.adsabs.harvard.edu/abs/2019A&A...627A.125M} {627, A125}

\bibitem[\protect\citeauthoryear{{Markwardt} \& {Swank}}{{Markwardt} \&
  {Swank}}{2005}]{Markwardt2005}
{Markwardt} C.~B.,  {Swank} J.~H.,  2005, The Astronomer's Telegram, \href
  {http://adsabs.harvard.edu/abs/2005ATel..495....1M} {495}

\bibitem[\protect\citeauthoryear{{Markwardt}, {Swank}, {Strohmayer}, {in 't
  Zand}  \& {Marshall}}{{Markwardt} et~al.}{2002}]{Markwardt2002}
{Markwardt} C.~B.,  {Swank} J.~H.,  {Strohmayer} T.~E.,  {in 't Zand} J.~J.~M.,
    {Marshall} F.~E.,  2002, \mn@doi [\apjl] {10.1086/342612}, \href
  {https://ui.adsabs.harvard.edu/abs/2002ApJ...575L..21M} {575, L21}

\bibitem[\protect\citeauthoryear{{Matsuoka} et~al.,}{{Matsuoka}
  et~al.}{2009}]{Matsuoka2009}
{Matsuoka} M.,  et~al., 2009, \mn@doi [\pasj] {10.1093/pasj/61.5.999}, \href
  {https://ui.adsabs.harvard.edu/abs/2009PASJ...61..999M} {61, 999}

\bibitem[\protect\citeauthoryear{{Mihara} et~al.,}{{Mihara}
  et~al.}{2011}]{Mihara2011}
{Mihara} T.,  et~al., 2011, \mn@doi [\pasj] {10.1093/pasj/63.sp3.S623}, \href
  {https://ui.adsabs.harvard.edu/abs/2011PASJ...63S.623M} {63, S623}

\bibitem[\protect\citeauthoryear{{Mukherjee}, {Bult}, {van der Klis}  \&
  {Bhattacharya}}{{Mukherjee} et~al.}{2015}]{Mukherjee2015}
{Mukherjee} D.,  {Bult} P.,  {van der Klis} M.,   {Bhattacharya} D.,  2015,
  \mn@doi [\mnras] {10.1093/mnras/stv1542}, \href
  {http://adsabs.harvard.edu/abs/2015MNRAS.452.3994M} {452, 3994}

\bibitem[\protect\citeauthoryear{{Negoro} et~al.,}{{Negoro}
  et~al.}{2017}]{Negoro2017}
{Negoro} H.,  et~al., 2017, The Astronomer's Telegram, \href
  {http://adsabs.harvard.edu/abs/2017ATel10821....1N} {10821}

\bibitem[\protect\citeauthoryear{{Ortolani}, {Barbuy}  \& {Bica}}{{Ortolani}
  et~al.}{1994}]{Ortolani1994}
{Ortolani} S.,  {Barbuy} B.,   {Bica} E.,  1994, \aaps, \href
  {http://adsabs.harvard.edu/abs/1994A%26AS..108..653O} {108, 653}

\bibitem[\protect\citeauthoryear{{Papitto}, {Di Salvo}, {D'A{\`i}}, {Iaria},
  {Burderi}, {Riggio}, {Menna}  \& {Robba}}{{Papitto}
  et~al.}{2009}]{Papitto2009}
{Papitto} A.,  {Di Salvo} T.,  {D'A{\`i}} A.,  {Iaria} R.,  {Burderi} L.,
  {Riggio} A.,  {Menna} M.~T.,   {Robba} N.~R.,  2009, \mn@doi [\aap]
  {10.1051/0004-6361:200811401}, \href
  {http://adsabs.harvard.edu/abs/2009A%26A...493L..39P} {493, L39}

\bibitem[\protect\citeauthoryear{{Papitto}, {Riggio}, {di Salvo}, {Burderi},
  {D'A{\`\i}}, {Iaria}, {Bozzo}  \& {Menna}}{{Papitto}
  et~al.}{2010}]{Papitto2010}
{Papitto} A.,  {Riggio} A.,  {di Salvo} T.,  {Burderi} L.,  {D'A{\`\i}} A.,
  {Iaria} R.,  {Bozzo} E.,   {Menna} M.~T.,  2010, \mn@doi [\mnras]
  {10.1111/j.1365-2966.2010.17090.x}, \href
  {https://ui.adsabs.harvard.edu/abs/2010MNRAS.407.2575P} {407, 2575}

\bibitem[\protect\citeauthoryear{{Papitto} et~al.,}{{Papitto}
  et~al.}{2013a}]{Papitto2013}
{Papitto} A.,  et~al., 2013a, \mn@doi [\mnras] {10.1093/mnras/sts605}, \href
  {http://adsabs.harvard.edu/abs/2013MNRAS.429.3411P} {429, 3411}

\bibitem[\protect\citeauthoryear{{Papitto} et~al.,}{{Papitto}
  et~al.}{2013b}]{Papitto2013b}
{Papitto} A.,  et~al., 2013b, \mn@doi [\nat] {10.1038/nature12470}, \href
  {https://ui.adsabs.harvard.edu/abs/2013Natur.501..517P} {501, 517}

\bibitem[\protect\citeauthoryear{{Patruno} \& {Watts}}{{Patruno} \&
  {Watts}}{2012}]{Patruno2012}
{Patruno} A.,  {Watts} A.~L.,  2012, preprint, \href
  {http://adsabs.harvard.edu/abs/2012arXiv1206.2727P} {} (\mn@eprint {arXiv}
  {1206.2727})

\bibitem[\protect\citeauthoryear{{Patruno}, {Altamirano}, {Hessels}, {Casella},
  {Wijnands}  \& {van der Klis}}{{Patruno} et~al.}{2009}]{Patruno2009}
{Patruno} A.,  {Altamirano} D.,  {Hessels} J.~W.~T.,  {Casella} P.,  {Wijnands}
  R.,   {van der Klis} M.,  2009, \mn@doi [\apj]
  {10.1088/0004-637X/690/2/1856}, \href
  {http://adsabs.harvard.edu/abs/2009ApJ...690.1856P} {690, 1856}

\bibitem[\protect\citeauthoryear{{Patruno} et~al.,}{{Patruno}
  et~al.}{2010}]{Patruno2010}
{Patruno} A.,  et~al., 2010, The Astronomer's Telegram, \href
  {http://adsabs.harvard.edu/abs/2010ATel.2407....1P} {2407}

\bibitem[\protect\citeauthoryear{{Pintore} et~al.,}{{Pintore}
  et~al.}{2014}]{Pintore2014}
{Pintore} F.,  et~al., 2014, \mn@doi [\mnras] {10.1093/mnras/stu2001}, \href
  {https://ui.adsabs.harvard.edu/abs/2014MNRAS.445.3745P} {445, 3745}

\bibitem[\protect\citeauthoryear{{Pintore} et~al.,}{{Pintore}
  et~al.}{2016}]{Pintore}
{Pintore} F.,  et~al., 2016, \mn@doi [\mnras] {10.1093/mnras/stw176}, \href
  {http://adsabs.harvard.edu/abs/2016MNRAS.457.2988P} {457, 2988}

\bibitem[\protect\citeauthoryear{{Pintore} et~al.,}{{Pintore}
  et~al.}{2018}]{Pintore2018}
{Pintore} F.,  et~al., 2018, \mn@doi [\mnras] {10.1093/mnras/sty1735}, \href
  {http://adsabs.harvard.edu/abs/2018MNRAS.479.4084P} {479, 4084}

\bibitem[\protect\citeauthoryear{{Pooley} et~al.,}{{Pooley}
  et~al.}{2002}]{Pooley2002}
{Pooley} D.,  et~al., 2002, \mn@doi [\apj] {10.1086/340498}, \href
  {http://adsabs.harvard.edu/abs/2002ApJ...573..184P} {573, 184}

\bibitem[\protect\citeauthoryear{{Pringle} \& {Rees}}{{Pringle} \&
  {Rees}}{1972}]{Pringle1972}
{Pringle} J.~E.,  {Rees} M.~J.,  1972, \aap, \href
  {https://ui.adsabs.harvard.edu/abs/1972A&A....21....1P} {21, 1}

\bibitem[\protect\citeauthoryear{{Riggio}, {Burderi}, {di Salvo}, {Papitto},
  {D'A{\`i}}, {Iaria}  \& {Menna}}{{Riggio} et~al.}{2011}]{Riggio2011}
{Riggio} A.,  {Burderi} L.,  {di Salvo} T.,  {Papitto} A.,  {D'A{\`i}} A.,
  {Iaria} R.,   {Menna} M.~T.,  2011, \mn@doi [\aap]
  {10.1051/0004-6361/201014883}, \href
  {https://ui.adsabs.harvard.edu/abs/2011A%26A...531A.140R} {531, A140}

\bibitem[\protect\citeauthoryear{{Sanna} et~al.,}{{Sanna}
  et~al.}{2016}]{Sanna2016}
{Sanna} A.,  et~al., 2016, \mn@doi [\mnras] {10.1093/mnras/stw740}, \href
  {http://adsabs.harvard.edu/abs/2016MNRAS.459.1340S} {459, 1340}

\bibitem[\protect\citeauthoryear{{Sanna} et~al.,}{{Sanna}
  et~al.}{2018a}]{Sanna2018a}
{Sanna} A.,  et~al., 2018a, \mn@doi [\mnras] {10.1093/mnras/sty2316}, \href
  {https://ui.adsabs.harvard.edu/abs/2018MNRAS.481.1658S} {481, 1658}

\bibitem[\protect\citeauthoryear{{Sanna} et~al.,}{{Sanna}
  et~al.}{2018b}]{Sanna2018b}
{Sanna} A.,  et~al., 2018b, \mn@doi [\aap] {10.1051/0004-6361/201732262}, \href
  {http://adsabs.harvard.edu/abs/2018A%26A...610L...2S} {610, L2}

\bibitem[\protect\citeauthoryear{{Sanna} et~al.,}{{Sanna}
  et~al.}{2018c}]{Sanna2018d}
{Sanna} A.,  et~al., 2018c, \mn@doi [\aap] {10.1051/0004-6361/201833205}, \href
  {https://ui.adsabs.harvard.edu/abs/2018A&A...616L..17S} {616, L17}

\bibitem[\protect\citeauthoryear{{Sanna} et~al.,}{{Sanna}
  et~al.}{2018d}]{Sanna2018e}
{Sanna} A.,  et~al., 2018d, \mn@doi [\aap] {10.1051/0004-6361/201834160}, \href
  {https://ui.adsabs.harvard.edu/abs/2018A&A...617L...8S} {617, L8}

\bibitem[\protect\citeauthoryear{{Sharma}, {Jain}  \& {Dutta}}{{Sharma}
  et~al.}{2019}]{Sharma2019}
{Sharma} R.,  {Jain} C.,   {Dutta} A.,  2019, \mn@doi [\mnras]
  {10.1093/mnras/sty2808}, \href
  {https://ui.adsabs.harvard.edu/abs/2019MNRAS.482.1634S} {482, 1634}

\bibitem[\protect\citeauthoryear{{Singh} et~al.,}{{Singh}
  et~al.}{2014}]{Singh2014}
{Singh} K.~P.,  et~al., 2014, in Space Telescopes and Instrumentation 2014:
  Ultraviolet to Gamma Ray. p. 91441S, \mn@doi{10.1117/12.2062667}

\bibitem[\protect\citeauthoryear{{Singh} et~al.,}{{Singh}
  et~al.}{2016}]{Singh2016}
{Singh} K.~P.,  et~al., 2016, in Space Telescopes and Instrumentation 2016:
  Ultraviolet to Gamma Ray. p. 99051E, \mn@doi{10.1117/12.2235309}

\bibitem[\protect\citeauthoryear{Singh et~al.,}{Singh et~al.}{2017}]{Singh2017}
Singh K.~P.,  et~al., 2017, \mn@doi [Journal of Astrophysics and Astronomy]
  {10.1007/s12036-017-9448-7}, 38, 29

\bibitem[\protect\citeauthoryear{{Sreehari}, {Ravishankar}, {Iyer}, {Agrawal},
  {Katoch}, {Mandal}  \& {Nand i}}{{Sreehari} et~al.}{2019}]{Sreehari2019}
{Sreehari} H.,  {Ravishankar} B.~T.,  {Iyer} N.,  {Agrawal} V.~K.,  {Katoch}
  T.~B.,  {Mandal} S.,   {Nand i} A.,  2019, \mn@doi [\mnras]
  {10.1093/mnras/stz1327}, \href
  {https://ui.adsabs.harvard.edu/abs/2019MNRAS.487..928S} {487, 928}

\bibitem[\protect\citeauthoryear{{Str{\"u}der} et~al.,}{{Str{\"u}der}
  et~al.}{2001}]{Struder2001}
{Str{\"u}der} L.,  et~al., 2001, \mn@doi [\aap] {10.1051/0004-6361:20000066},
  \href {https://ui.adsabs.harvard.edu/abs/2001A&A...365L..18S} {365, L18}

\bibitem[\protect\citeauthoryear{{Turner} et~al.,}{{Turner}
  et~al.}{2001}]{Turner2001}
{Turner} M.~J.~L.,  et~al., 2001, \mn@doi [\aap] {10.1051/0004-6361:20000087},
  \href {https://ui.adsabs.harvard.edu/abs/2001A&A...365L..27T} {365, L27}

\bibitem[\protect\citeauthoryear{{Valenti}, {Ferraro}  \& {Origlia}}{{Valenti}
  et~al.}{2007}]{Valenti2007}
{Valenti} E.,  {Ferraro} F.~R.,   {Origlia} L.,  2007, \mn@doi [\aj]
  {10.1086/511271}, \href {http://adsabs.harvard.edu/abs/2007AJ....133.1287V}
  {133, 1287}

\bibitem[\protect\citeauthoryear{{Verner}, {Ferland}, {Korista}  \&
  {Yakovlev}}{{Verner} et~al.}{1996}]{Verner}
{Verner} D.~A.,  {Ferland} G.~J.,  {Korista} K.~T.,   {Yakovlev} D.~G.,  1996,
  \mn@doi [\apj] {10.1086/177435}, \href
  {http://adsabs.harvard.edu/abs/1996ApJ...465..487V} {465, 487}

\bibitem[\protect\citeauthoryear{{Wijnands} \& {van der Klis}}{{Wijnands} \&
  {van der Klis}}{1998}]{Wijnands1998}
{Wijnands} R.,  {van der Klis} M.,  1998, \mn@doi [\nat] {10.1038/28557}, \href
  {https://ui.adsabs.harvard.edu/abs/1998Natur.394..344W} {394, 344}

\bibitem[\protect\citeauthoryear{{Wilkinson}, {Patruno}, {Watts}  \&
  {Uttley}}{{Wilkinson} et~al.}{2011}]{Wilkinson2011}
{Wilkinson} T.,  {Patruno} A.,  {Watts} A.,   {Uttley} P.,  2011, \mn@doi
  [\mnras] {10.1111/j.1365-2966.2010.17532.x}, \href
  {http://adsabs.harvard.edu/abs/2011MNRAS.410.1513W} {410, 1513}

\bibitem[\protect\citeauthoryear{{Wilms}, {Allen}  \& {McCray}}{{Wilms}
  et~al.}{2000}]{Wilms}
{Wilms} J.,  {Allen} A.,   {McCray} R.,  2000, \mn@doi [\apj] {10.1086/317016},
  \href {http://adsabs.harvard.edu/abs/2000ApJ...542..914W} {542, 914}

\bibitem[\protect\citeauthoryear{{Wu}, {Ding}, {Li}, {Chen}  \& {Qu}}{{Wu}
  et~al.}{2018}]{Wu2018}
{Wu} Z.,  {Ding} G.,  {Li} Z.,  {Chen} Y.,   {Qu} J.,  2018, \mn@doi [\apss]
  {10.1007/s10509-018-3367-1}, \href
  {https://ui.adsabs.harvard.edu/abs/2018Ap&SS.363..146W} {363, 146}

\bibitem[\protect\citeauthoryear{{Yadav} et~al.,}{{Yadav}
  et~al.}{2016}]{Yadav2016}
{Yadav} J.~S.,  et~al., 2016, in Space Telescopes and Instrumentation 2016:
  Ultraviolet to Gamma Ray. p. 99051D, \mn@doi{10.1117/12.2231857}

\bibitem[\protect\citeauthoryear{{Zdziarski}, {Johnson}  \&
  {Magdziarz}}{{Zdziarski} et~al.}{1996}]{zdziarski}
{Zdziarski} A.~A.,  {Johnson} W.~N.,   {Magdziarz} P.,  1996, \mn@doi [\mnras]
  {10.1093/mnras/283.1.193}, \href
  {http://adsabs.harvard.edu/abs/1996MNRAS.283..193Z} {283, 193}

\bibitem[\protect\citeauthoryear{{{\.Z}ycki}, {Done}  \& {Smith}}{{{\.Z}ycki}
  et~al.}{1999}]{zycki}
{{\.Z}ycki} P.~T.,  {Done} C.,   {Smith} D.~A.,  1999, \mn@doi [\mnras]
  {10.1046/j.1365-8711.1999.02885.x}, \href
  {http://adsabs.harvard.edu/abs/1999MNRAS.309..561Z} {309, 561}

\bibitem[\protect\citeauthoryear{{in 't Zand} et~al.,}{{in 't Zand}
  et~al.}{1999}]{intZand1999}
{in 't Zand} J.~J.~M.,  et~al., 1999, \aap, \href
  {http://adsabs.harvard.edu/abs/1999A%26A...345..100I} {345, 100}

\bibitem[\protect\citeauthoryear{{in't Zand}, {van Kerkwijk}, {Pooley},
  {Verbunt}, {Wijnands}  \& {Lewin}}{{in't Zand} et~al.}{2001}]{intZand2001}
{in't Zand} J.~J.~M.,  {van Kerkwijk} M.~H.,  {Pooley} D.,  {Verbunt} F.,
  {Wijnands} R.,   {Lewin} W.~H.~G.,  2001, \mn@doi [\apjl] {10.1086/338361},
  \href {http://adsabs.harvard.edu/abs/2001ApJ...563L..41I} {563, L41}

\makeatother
\end{thebibliography}
\end{document}